\documentclass[longauth]{aa}  
\usepackage{xfrac}
\usepackage{lscape}
\usepackage{placeins}
\usepackage{graphicx}
\usepackage{txfonts}
\usepackage{hyperref}
\hypersetup{
    colorlinks=true,
    linkcolor=blue,
    filecolor=blue,
    urlcolor=blue,
    pdftitle={LEWIS III -- UDG32},
    citecolor=blue,
    }
\usepackage{natbib}
\bibpunct{(}{)}{;}{a}{}{,} 

\newcommand\T{\rule{0pt}{2.6ex}}       
\newcommand\B{\rule[-1.2ex]{0pt}{0pt}} 

\begin{document}

   \title{Looking into the faintEst WIth MUSE (LEWIS): Exploring the nature of ultra-diffuse galaxies in the Hydra~I cluster}
   \subtitle{III. Untangling UDG~32 from the stripped filaments of NGC~3314A  with multi-wavelength data}
   \titlerunning{LEWIS III. Untangling UDG~32 from the stripped filaments of NGC~3314a in the Hydra~I cluster}
   \authorrunning{J. Hartke et al.}


   \author{J. Hartke\inst{\ref{FINCA},\ref{UTU}}\fnmsep\thanks{\email{johanna.hartke@utu.fi}} 
           \and E. Iodice\inst{\ref{INAF-N}} 
           \and M. Gullieuszik\inst{\ref{INAF-P}} 
           \and M. Mirabile\inst{\ref{INAF-T},\ref{ESO-G},\ref{GSSI}} 
           \and C. Buttitta\inst{\ref{INAF-N}} 
           \and G. Doll\inst{\ref{INAF-N},\ref{UniNA}} 
           \and G. D'Ago\inst{\ref{IoA}} 
           \and C.~C. de la Casa\inst{\ref{CSIC}} 
           \and K.~M.~Hess\inst{\ref{Chalmers},\ref{ASTRON}} 
           \and R. Kotulla\inst{\ref{UW–Madison}} 
           \and B. Poggianti\inst{\ref{INAF-P}} 
           \and M. Arnaboldi\inst{\ref{ESO-G}} 
           \and M. Cantiello\inst{\ref{INAF-T}} 
           \and E.~M. Corsini\inst{\ref{UniPa},\ref{INAF-P}} 
          \and J. Falc{\'o}n-Barroso \inst{\ref{IAC},\ref{ULL}} 
          \and D.~A.~Forbes \inst{\ref{Swinburne}} 
          \and M. Hilker\inst{\ref{ESO-G}} 
          \and S. Mieske \inst{\ref{ESO-CL}} 
          \and M. Rejkuba \inst{\ref{ESO-G}} 
          \and M. Spavone \inst{\ref{INAF-N}} 
          \and C. Spiniello \inst{\ref{Oxford},\ref{INAF-N}} 
          }

    \institute{
              Finnish Centre for Astronomy with ESO,
              (FINCA), University of Turku, 20014 Turku, Finland \label{FINCA}             
              \and
              Tuorla Observatory, Department of Physics and Astronomy, University of Turku, 20014 Turku, Finland \label{UTU}
              \and
              INAF -- Osservatorio Astronomico di Capodimonte, Salita Moiariello 16, 80131 Napoli, Italy \label{INAF-N}
              \and
              INAF -- Osservatorio Astronomico di Padova, Vicolo dell'Osservatorio 5, 35122 Padova, Italy \label{INAF-P}
              \and
              INAF -- Osservatorio Astronomico di Abruzzo, Via Maggini, 64100 Teramo, Italy \label{INAF-T}
              \and
              European Southern Observatory, Karl-Schwarzschild-Straße 2, 85748 Garching bei München, Germany \label{ESO-G}
              \and
              Gran Sasso Science Institute, Viale Francesco Crispi 7, I-67100 L’Aquila, Italy \label{GSSI}
              \and
              University of Naples “Federico II”, C.U. Monte Sant’Angelo, Via
              Cinthia, 80126, Naples, Italy \label{UniNA}
              \and 
              Institute of Astronomy, University of Cambridge, Madingley Road, Cambridge CB3 0HA, UK \label{IoA}
              \and
              Instituto de Astrofísica de Andalucía (CSIC), Glorieta de la Astronomía s/n, 18008 Granada, Spain \label{CSIC}
              \and
              Department of Space, Earth and Environment, Chalmers University of Technology, Onsala Space Observatory, 43992 Onsala, Sweden \label{Chalmers}
              \and
              ASTRON, the Netherlands Institute for Radio Astronomy, Postbus 2, 7990 AA, Dwingeloo, The Netherlands \label{ASTRON}
              \and
              Department of Astronomy, University of Wisconsin-Madison, 475 N Charter St, Madison, WI, 53706, USA \label{UW–Madison}
              \and
              Dipartimento di Fisica e Astronomia `G. Galilei', Universit\`a di Padova, Vicolo dell'Osservatorio 3, I-35122 Padova, Italy \label{UniPa}
              \and
              Instituto de Astrof\'isica de Canarias, Calle V\'ia L\'actea s/n, E-38205. La Laguna, Tenerife, Spain \label{IAC}
              \and
              Departamento de Astrof\'isica. Universidad de La Laguna, Av. del Astrof\'isico Francisco S\'anchez s/n, E-38206, La Laguna, Tenerife, Spain \label{ULL}
              \and
              Centre for Astrophysics \& Supercomputing, Swinburne University of Technology, Hawthorn VIC 3122, Australia \label{Swinburne}
              \and 
              European Southern Observatory,
              Alonso de Cordova 3107, Vitacura,
              Casilla 19001, Santiago de Chile, Chile \label{ESO-CL}
              \and
              Sub-Department of Astrophysics, Department of Physics, University of Oxford, Denys Wilkinson Building, Keble Road, Oxford OX1 3RH, United Kingdom \label{Oxford}
              }
              
   \date{\today}

 
  \abstract
   {UDG~32 is an ultra-diffuse galaxy (UDG) candidate in the Hydra~I cluster that was discovered in the extended network of stellar filaments of the jellyfish galaxy NGC~3314A. This jellyfish galaxy is affected by ram pressure stripping and it is hypothesised that UDG~32 may have formed from this stripped material.}
   {The aim of this paper is to address whether UDG~32 can be associated with the stripped material of NGC~3314A and to constrain its formation scenario in relation to its environment.}
   {We use new integral-field spectroscopic data from the MUSE large programme `LEWIS' in conjunction with deep multi-band photometry to constrain the kinematics of UDG~32 via spectral fitting and its stellar population properties with spectral energy distribution fitting.}
   {The new MUSE data allow us to reveal that the stripped material from NGC~3314A, traced by emission lines such as H$\alpha$, extends much further from its parent galaxy than previously known, completely overlapping with UDG~32 in projection, and with ram pressure induced star formation. We determine the line-of-sight velocity of UDG~32 to be $v_\mathrm{LOS} = 3080\pm120\,\mathrm{km}\,\mathrm{s}^{-1}$ and confirm that UDG~32 is part of the same kinematic structure as NGC~3314A, the Hydra~I cluster south-east subgroup. By fitting the UV and optical spectral energy distribution obtained from deep multi-band photometry, we constrain the stellar population properties of UDG~32. We determine its mass-weighted age to be $7.7^{+2.9}_{-2.8}\,\mathrm{Gyr}$ and its metallicity to be $[\mathrm{M/H}] = 0.07^{+0.19}_{-0.32}$ dex. We confirm the presence of two globular clusters (GCs) in the MUSE field of view, bound to the Hydra~I cluster rather than to UDG~32, making them part of the Hydra~I intracluster GC population.}
   {The metal-rich and intermediate-age nature of UDG~32 points towards its formation from pre-enriched material in the south-east group of the Hydra~I cluster that was liberated from a more massive galaxy via tidal or ram-pressure stripping, but we cannot establish a direct link to the ram-pressure stripped material from NGC~3314A.} 


   \maketitle
%

\section{Introduction}
Ultra-diffuse galaxies (UDGs) are fascinating objects characterised by their low surface brightness ($\mu_{0,g} \geq 24\,\mathrm{mag}\,\mathrm{arcsec}^{-2}$) and comparatively large sizes ($R_\mathrm{eff} > 1.5\,\mathrm{kpc}$). The term UDG and definition according to structural parameters was introduced by \citet{2015ApJ...798L..45V} following the discovery of a large population of low-surface brightness (LSB) galaxies in the Coma Cluster. 
Since then, several imaging campaigns have revealed that UDGs span a large range of photometric and structural properties, extending and overlapping with the parameter space of LSBs. 
The definition based on a strict cut in structural parameters \citep{2015ApJ...798L..45V} is somewhat arbitrary \citep[e.g.,][]{2022ApJ...926...92V}. UDGs can be interpreted in the context of a continuation of the LSB galaxy sequence on the mass -- effective radius diagnostic diagram towards lower (stellar) masses, rather than as a completely distinct population \citep{2018RNAAS...2...43C, 2024ApJS..271...52Z, 2024MNRAS.534.1729C}. However, UDGs may have distinct other properties such as their globular cluster (GC) content \citep[e.g.,][]{2018ApJ...862...82L, Forbes2020b, 2025MNRAS.536.1217F} or their halo mass \citep[e.g.,][]{2018ApJ...856L..31T, Toloba2023, Gannon2022, Forbes2024}.

Observationally, UDGs can be divided into a population of ``red'' UDGs, which are predominantly found in cluster and group environments, and ``blue'' UDGs that are predominantly found in less dense environments such as cluster outskirts and the field \citep[e.g.,][]{2017MNRAS.468..703R, 2019MNRAS.488.2143P, 2021A&A...654A.105M}. This diversity points towards different formation scenarios for UDGs, mainly depending on their environment, as summarised in \citet{2023A&A...679A..69I}. 

To constrain these mechanisms during UDG formation and evolution, it is imperative to measure their stellar kinematics and stellar population properties from spectroscopy. However, considering the low-surface brightness nature of UDGs, this is a telescope-time-intensive effort. The LEWIS (Looking into the faintest with MUSE) programme\footnote{\url{https://sites.google.com/inaf.it/lewis/home}} has obtained the first homogeneous integral-field spectroscopic survey of 30 UDGs and low-surface brightness galaxies in the Hydra~I Cluster \citep{2023A&A...679A..69I}, following their discovery from deep optical imaging \citep{2020A&A...642A..48I, 2022A&A...665A.105L}. 

In this paper, we focus on the faintest UDG in the LEWIS sample with a central surface brightness of $\mu_{0,g} = 26\pm1\,\mathrm{mag}\,\mathrm{arcsec}^{-2}$,  designated UDG~32, and discovered by \citet{2021A&A...652L..11I}. What makes this object unique is that it lies in the stellar filaments of NGC~3314A, at a projected distance of 161\farcs5 from the centre of the galaxy.
If assuming that UDG~32 lies at the mean distance of the Hydra cluster ($51\pm6$~Mpc), its size is $R_\mathrm{eff} = 3.8\pm1.0$~kpc\footnote{If we instead assume that it is at the distance of NGC~3314A \citep[$\sim37.3$~Mpc,][]{2003Ap&SS.285..197C}, the physical effective radius is $R_\mathrm{eff} = 2.5\pm1.0$~kpc, still consistent with the conservative UDG definition of \citet{2015ApJ...798L..45V}.}. Its colour, $g-r = 0.54\pm0.14\,\mathrm{mag}$  and absolute $r$-band magnitude $M_r = -14.65\,\mathrm{mag}$ place it onto the colour-magnitude relation of dwarf galaxies in the Hydra~I cluster \citep{2008A&A...486..697M, 2022A&A...659A..92L}.
While its structural parameters do not set it apart from other UDGs in the Hydra~I cluster \citep{2022A&A...665A.105L}, UDG~32 is the only UDG candidate to date that has been detected in the vicinity of the filaments of a jellyfish galaxy so far \citep{2021A&A...652L..11I}. 

\begin{figure*}
    \centering
    \includegraphics[width=18cm]{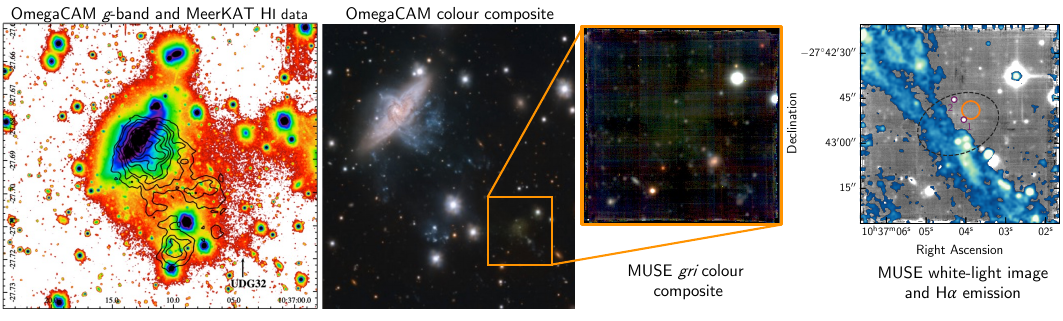}
    \caption{Multi-wavelength view of UDG 32 and its environment. From \textit{left} to \textit{right}: VST OmegaCAM $g$-band image \citep{2021A&A...652L..11I} of UDG 32 and NGC~3314A/B with black contours tracing the \ion{H}{i} emission detected by the MeerKAT survey \citep{2022A&A...668A.184H}, VST OmegaCAM colour-composite image \citep{2021A&A...652L..11I} with the location of UDG~32 highlighted by the orange box, colour-composite derived from the MUSE data presented in this paper, and MUSE white-light image with the H$\alpha$ emission overplotted by the blue colour map. The dashed ellipse denotes the 1~$R_\mathrm{eff}$ isophote of UDG~32 derived by \citet{2021A&A...652L..11I} and the purple circles the two intracluster GCs that were spectroscopically confirmed in this work. The orange circle denotes the aperture from which we extract the spectrum to derive the LOS velocity of UDG~32. Coordinates in this and all following figures are in J2000.0, and North is up and East to left.}
    \label{fig:overview}
\end{figure*} 

Jellyfish galaxies \citep[e.g.,][]{2007MNRAS.376..157C, 2010MNRAS.408.1417S, 2014ApJ...781L..40E, 2017ApJ...844...48P} are the result of gas-rich galaxies interacting with the hot intracluster medium. This is due to ram pressure stripping (RPS), i.e. the removal of cold gas and interstellar medium as galaxies move through the hot intracluster medium \citep{1972ApJ...176....1G}. 
Depending on the location of jellyfish galaxies in the phase space of the cluster with which they are interacting, stripped tails start appearing after the galaxies fall into the cluster for the first time and remain visible for a few hundred Myr past their pericentric passage \citep{2022ApJ...934...86S}. \citet{2024MNRAS.533..341S} estimated a total visibility lifetime of the stripped tails of $\Delta t=1.73\pm0.48$ Gyr based on a large sample of optically selected RPS galaxies \citep{2016AJ....151...78P, 2022ApJ...927...91V}. In simulations such as IllustrisTNG, RPS can act on galaxies for longer time scales ($1.5-8$ Gyr), but during the peak stripping period, when jellyfish galaxies are at host-centric distances comparable to that probed by \citet[][$\sim 0.2-2R_{200}$]{2024MNRAS.533..341S}, the stripping times are found to last $\lesssim 2$ Gyr \citep{2023MNRAS.524.3502R}, in agreement with observations. 

The co-location of the UDG candidate and the jellyfish tails is especially interesting in the context of several UDG formation scenarios. The typical (stellar) masses of RPS gas clumps span several orders of magnitude, ranging from $~\sim 10^{3.5}$ to $\sim 10^{8}\,M_\odot$ \citep[][see also Table~\ref{tab:scenarios}]{2019MNRAS.482.4466P, 2024A&A...682A.162W}, with the massive end overlapping with the mass distribution of observed UDGs. Hence, \citet{2019MNRAS.482.4466P} suggested that these clumps may be viable progenitors of dark matter (DM) free UDGs. This formation channel has also recently been explored in hydrodynamical cosmological simulations \citep{2024ApJ...969...24L}. 

Another formation scenario for DM-free dwarfs are tidal interactions between galaxies as proposed by \citet{2014MNRAS.440.1458D} following the observation of tidal dwarf galaxies (TDGs) of which some have structural parameters in the UDG domain. In an idealised simulation setup, \citet{2024A&A...687A.105I} combined the scenarios of tidal interaction and RPS and showed that ``wet'' galaxy mergers in clusters can produce several DM-free dwarf galaxies, some of them with UDG-like structural parameters. 
UDG~32 may thus be a hitherto unexplored laboratory of galaxy formation in the context of the above scenarios.

In the two left panels of Fig.~\ref{fig:overview}, we highlight the location of UDG~32 in the tails of the disturbed spiral galaxy NGC~3314A that is seen in projection on top of NGC~3314B. Based on their heliocentric velocities, NGC~3314A is in the foreground ($cz=2795\,\mathrm{km}\,\mathrm{s}^{-1}$), while NGC~3314B is in the background ($cz=4665\,\mathrm{km}\,\mathrm{s}^{-1}$). Both galaxies have \ion{H}{i} emission, with NGC~3314A exhibiting a complex velocity field with two distinct tails \citep{2022A&A...668A.184H}. 
The galaxy is part of a late-type galaxy group around NGC~3312 that interacts with the Hydra~I cluster \citep{2003Ap&SS.285..197C, 2021ApJ...915...70W, 2022A&A...668A.184H, 2024A&A...689A.306S}. \citet{2022A&A...668A.184H} estimate NGC~3314A to be $\sim 800\,\mathrm{kpc}$ and NGC~3312 at $\sim 550\,\mathrm{kpc}$ from the cluster centre and propose that the foreground group is moving towards us having already passed the cluster pericentre. 
Both the Wallaby \citep{2021ApJ...915...70W} and MeerKAT \citep{2022A&A...668A.184H} \ion{H}{i} surveys conclude that the displaced material in the tails is due to RPS and based on its morphology and gas content, NGC~3314A has been classified as a jellyfish galaxy \citep{2024MNRAS.528.1125K, 2024MNRAS.532..270G}. 

Based on photometry alone, it is impossible to rule out whether UDG~32 is a foreground or background galaxy with respect to NGC~3314A. Therefore, in this paper, we address this open question with new integral-field spectroscopic data. Complemented with ancillary multi-band photometry, we can determine the location of UDG~32 in the Hydra~I cluster phase space, assess its GC and stellar population content, and distinguish between different formation scenarios. If confirmed, UDG~32  would be the first discovery of a UDG formed from RPS clumps.

This paper is organised as follows: in Sect.~\ref{sect:data} we describe the new MUSE data covering UDG~32 and synergies with archival imaging data. In Sect.~\ref{sect:emission} we describe our discovery of the gaseous filaments of NGC~3314a in projection of UDG~32 and characterise them. In Sect.~\ref{sect:properties} and Sect.~\ref{sect:SSP} we derive kinematic and stellar population properties of UDG~32 from integral-field spectroscopy and its spectral energy distribution. In Sect.~\ref{sect:discussion} we discuss the formation scenarios for UDG~32 in light of our new results. 

\section{Data}
\label{sect:data}
UDG~32 has been observed with the Multi Unit Spectroscopic Explorer  \citep[MUSE,][]{2010SPIE.7735E..08B} at the ESO \textit{Very Large Telescope} in the scope of the LEWIS large programme (PI: E. Iodice, ID: 108.222P) with the standard LEWIS observing strategy \citep{2023A&A...679A..69I}. Seven observing blocks (OBs) with three exposures each were observed. In two OBs, the individual exposures were 750\,s long, while in the remaining five OBs the exposures were 900\,s long. The final combined cube has a seeing-weighted exposure time of 4.3\,h.

The presence of notable emission features in the quick-reduced data motivated us to develop an improved data reduction workflow with the goal to minimise spurious absorption features due to the emission being misinterpreted as sky. The workflow differs from the standard LEWIS data reduction procedures \citep{2023A&A...679A..69I} in the use of more conservative sky masks and sigma-clipped autocalibration factors, and is described in detail in Appendix~\ref{app:datared}.

The reconstructed $gri$ colour composite and white-light image (obtained by summing over the entire spectral axis) resulting from the improved data reduction workflow are shown in the two rightmost panels of Fig.~\ref{fig:overview}. To first order, the colours derived from VST OmegaCAM and MUSE data agree well. In the rightmost panel, we also illustrate the unprecedented extension of the H$\alpha$ emission at $z\approx0.01$ with blue contours that we discuss in Sect.~\ref{sect:emission}. This extended emission covering approximately half of UDG~32 within one effective radius limits the area over which we can stack for an absorption-line spectroscopic analysis. As a consequence, the signal-to-noise ratio (SNR) of the stacked spectra is too low for a stellar population analysis via absorption-line fitting, as we discuss in Sects.~\ref{sect:properties} and \ref{sect:SSP}. We therefore make use of ancillary optical and UV imaging from DECam and GALEX that we describe in Appendix~\ref{app:ancdata} in addition to the VST broad-band photometry shown in Fig.~\ref{fig:overview}. This allows us to constrain the stellar population properties of UDG~32 through an analysis of its spectral energy distribution (SED). 

\section{Revealing stripped material from NGC 3314A at unprecedented distances}
\label{sect:emission}
\begin{figure*}
    \includegraphics[width=18cm]{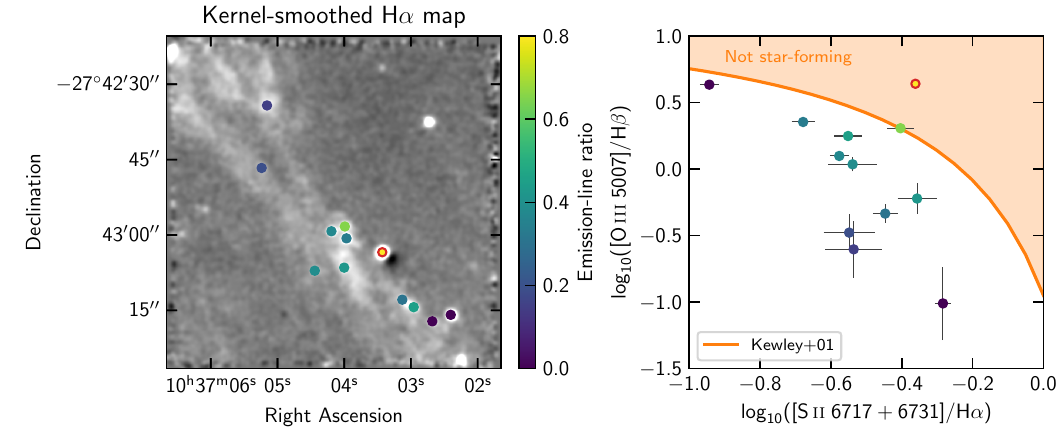}
    \caption{\textit{Left:} Knots colour-coded by their emission-line ratios on the kernel-smoothed H$\alpha$ map. In both panels, the point circled in red denotes a knot whose spectrum is likely contaminated by a background galaxy (appearing as a black feature on the kernel-smoothed H$\alpha$ map). \textit{Right:} BPT diagram of knots in the stripped arms of NGC 3314A with the same colour-coding. The orange line denotes the theoretical demarcation between starbursts and galaxies hosting active galactic nuclei \citep{2001ApJ...556..121K}.}
    \label{fig:BPT}
\end{figure*}
To date, \ion{H}{i} and H$\alpha$ emission (via narrow-band imaging) has been detected up to $\sim 2\arcmin$ from NGC~3314A in the direction of UDG~32, but no emission has been detected in association with, or beyond the UDG \citep{2021ApJ...915...70W, 2022A&A...668A.184H}. It was, therefore, unexpected to find clear signatures of several emission lines (H$\alpha$, H$\beta$, [\ion{O}{i}], [\ion{O}{iii}], [\ion{N}{ii}], [\ion{S}{ii}]) at redshift $z\sim0.01$ already in the quick-reduced data. We construct a map centred on the peak H$\alpha$ emission ($\lambda_\mathrm{c,obs} = 6630\,\r{A}$, $\Delta\lambda_c = \pm10\,\r{A}$) and subtract the continuum in two adjacent wavelength ranges ($\lambda_\mathrm{l,obs} = 6590\,\r{A}$ and $\lambda_\mathrm{r,obs} = 6670\,\r{A}$). The resulting map, convolved with a Gaussian kernel, is shown as a blue overlay in the rightmost panel of Fig.~\ref{fig:overview} and in full in the left panel of Fig.~\ref{fig:BPT}. 

The geometry and location of the emission features already point towards a relation with the stripped material of NGC~3314A. Further evidence comes from the velocity structure: the velocity in the filament ranges from $v_\mathrm{los} = 3003 \pm 10\,\mathrm{km}\,\mathrm{s}^{-1}$ in the northeast to  $v_\mathrm{los} = 2997 \pm 9\,\mathrm{km}\,\mathrm{s}^{-1}$ in the southwest of the MUSE FoV. 
This is comparable to the \ion{H}{i} velocities of NGC~3314A ($v_\mathrm{los} = 2849\,\mathrm{km}\,\mathrm{s}^{-1}$) and its stripped material (ranging from $v_\mathrm{los} = 2700\,\mathrm{km}\,\mathrm{s}^{-1}$ to $3000\,\mathrm{km}\,\mathrm{s}^{-1}$) that were derived by \citet{2022A&A...668A.184H}. We have therefore revealed the stripped material of NGC~3314A to unprecedented distances ($> 3\arcmin$), reaching beyond UDG~32 in projection. 

To characterise the stripped material, we focus on candidate star-forming regions, as ram pressure can trigger star formation (SF) when the interstellar medium is compressed \citep[e.g.,][]{1986ApJ...301...57B}. 
We use \textsc{daofind} implemented in \textsc{photutils} \citep{larry_bradley_2023_7946442} to detect knots along the stripped arms of NGC~3314A on the kernel-smoothed H$\alpha$ map that we presented earlier in this section. The left panel of Fig.~\ref{fig:BPT} shows their detection on the H$\alpha$ map. To confirm our hypothesis that these are star-forming knots, we extract their spectra and that of the underlying filament in circular apertures ($r=0\farcs6$) and fit their emission lines and kinematics in terms of the line-of-sight (LOS) velocity and velocity dispersion using \textsc{pPXF} \citep{2004PASP..116..138C, Cappellari2023}.  We construct a \citet*[][BPT]{1981PASP...93....5B} diagram based on the fitted H$\alpha$, H$\beta$, [\ion{O}{iii}], and [\ion{S}{ii}] emission-line fluxes, as shown in the right panel of Fig.~\ref{fig:BPT}. The SNRs of the fitted lines range from $\sim 2$ to $\sim 50$ for forbidden lines and from $\sim 5$ to $\sim 250$ for the two Balmer lines. To simplify the visual identification of knots on both panels of the figure, we colour-code them by their emission-line ratios as introduced by \citet{2019MNRAS.485L..38D}. 

According to the theoretical demarcation line of \citet{2001ApJ...556..121K}, we classify 11/12 knots as star forming, in line with the scenario of ram-pressure triggered SF. The object above the line is in close proximity to a $z=0.77$ background galaxy (see Appendix~\ref{app:cont_galaxy}). Considering that it has a velocity similar to that of the other \ion{H}{ii} knots (cf. Fig.~\ref{fig:vel_distr} in the Appendix), we consider that its spectrum may have been contaminated by the background galaxy. 
We derive the abundances of the 11 confirmed star-forming knots in Appendix~\ref{app:abundances}. All of them have sub-solar oxygen abundances (Fig.~\ref{fig:O/H_distr}) and there is no trend with distance from UDG 32 or, in extension, with NGC~3314A that lies along the same transversal line. There is quite a large scatter in the extinction $A_{V}$, with knots in the northeast of UDG~32 (i.e. towards NGC~3314A) having extinction values that are comparable with that of the interspersed dusty arms of NGC~3314A \citep{2001AJ....122.1369K}, while those in the southeast having on average higher values, but with a large scatter (Fig.~\ref{fig:Av_distr}). 

\section{Properties of UDG~32 from integral-field spectroscopy}
\label{sect:properties}
Having characterised the stripped material from NGC~3314A in the previous section, we now turn to the characterisation of UDG~32 based on the new MUSE data from LEWIS  and focus on its kinematics (Sect.~\ref{ssect:LOSV}) and GC content (Sect.~\ref{sect:GCs}).

\subsection{Line-of-sight velocity determination and link with NGC~3314A}
\label{ssect:LOSV}
In \citet{2023A&A...679A..69I} and \citet{Buttitta2025}, we derive the LOS velocity of the LEWIS UDGs based on their redshifted absorption lines. In the case of UDG~32, this is complicated by the presence of the foreground filament described in the previous section, whose strong emission lines dominate the spectra in the $1~R_\mathrm{eff}$ aperture (see right panel in Fig.~\ref{fig:overview}). It is not possible to identify any absorption lines. Instead, we extract a spectrum in a $3\arcsec$-radius circular aperture that does neither overlap with the H$\alpha$ filaments, nor with the two candidate GCs, but is as close to the centre of UDG~32 as possible. This aperture is denoted by the orange circle in the rightmost panel of Fig.~\ref{fig:overview}. There is some residual emission from the diffuse interstellar medium (ISM) at redshift $z=0$ (see Appendix~\ref{app:discoveries} and Fig.~\ref{fig:ISM_z0} for details) that we mask before any further analysis. 
We fit the spectra using \textsc{pPXF} in conjunction with the E-MILES stellar library \citep{2011A&A...532A..95F, 2016MNRAS.463.3409V} as detailed in Appendix~\ref{app:kinematics}. As the SNR of the fitted spectrum is $\sim 3$, well below 15, we will not be able to recover velocity dispersions below the MUSE instrumental resolution \citep[cf. Sects.~5.1 and 5.2 in][]{2023A&A...679A..69I} and we therefore focus on obtaining the LOS velocity to address the open question of the location of UDG~32 in the Hydra~I cluster phase space. We derive a conservative LOS velocity estimate of $v_\mathrm{LOS} = 3080\pm120\,\mathrm{km}\,\mathrm{s}^{-1}$ from the full optical spectrum and also fit just the region around the H$\alpha$ absorption, resulting in $v_\mathrm{LOS} = 3085\pm38\,\mathrm{km}\,\mathrm{s}^{-1}$. The errors on the LOS velocities were derived via bootstrapping as detailed in Appendix~\ref{app:kinematics}.  

\begin{figure}
    \centering
    \includegraphics[width=8.8cm]{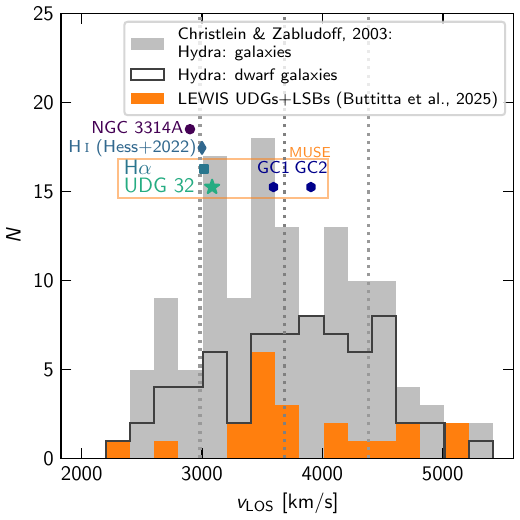}
    \caption{Velocity distribution of galaxies in the Hydra I cluster \citep[grey histogram,][]{2003Ap&SS.285..197C}, the LEWIS UDGs and LSBs \citep[orange histogram,]{Buttitta2025}, NGC~3314A \citep[purple circle,][]{2003MNRAS.339..652K}, and the ram-pressure stripped \ion{H}{i} gas \citep[teal diamond,][]{2022A&A...668A.184H}. Velocities derived from the LEWIS observations of UDG~32 are in the orange box: the H$\alpha$ arm velocity (square), the velocity of UDG~32 based on the short spectrum around H$\alpha$ (star), and the velocities of the two GCs (hexagons). The mean velocity of the Hydra~I cluster and its $1\sigma$-limits \citep{2021MNRAS.500.1323L} are denoted with vertical dotted lines.}
    \label{fig:vhist}
\end{figure}

Figure~\ref{fig:vhist} summarises our results and shows the kinematic proximity of UDG~32 and the filament traced by H$\alpha$ and \ion{H}{i} gas to NGC~3314A. While we stress that we did not determine a physical distance to UDG~32, this points to a relation between NGC~3314A, its stripped gas, and UDG~32. Based on their velocities, they all seem to belong to the same foreground velocity substructure in the south-east region of the Hydra~I cluster. Alongside the late-type NGC~3312 that also shows signs of RPS, NGC~3314A is one of the brightest galaxies of this group \citep{2003Ap&SS.285..197C, 2021ApJ...915...70W, 2022A&A...668A.184H, 2024A&A...689A.306S}. 

\subsection{Globular cluster content}
\label{sect:GCs}
GCs can provide independent constraints on the kinematics, stellar populations, and matter content of galaxies and also on their formation scenario. Recently formed galaxies, e.g. through tidal interaction \citep{2014MNRAS.440.1458D} or via RPS \citep{2019MNRAS.482.4466P}, are not expected to host GCs. In contrast, those with old stellar populations, e.g. so-called failed galaxies \citep{2015ApJ...798L..45V} or tidally heated `normal' dwarf galaxies \citep{2021ApJ...919...72J}, are expected to host several GCs. 

We spectroscopically confirm the presence of two GCs in the MUSE FoV, whose source extraction and fitting is described in Appendix~\ref{app:GCs}. Their LOS velocities are $v_\mathrm{LOS,GC1} = 3594\pm23\,\mathrm{km}\,\mathrm{s}^{-1}$ and $v_\mathrm{LOS,GC2} = 3905\pm26\,\mathrm{km}\,\mathrm{s}^{-1}$. As illustrated by the blue hexagons in Fig.~\ref{fig:vhist}, these velocities are several hundreds $\mathrm{km}\,\mathrm{s}^{-1}$ larger than that of UDG~32. Therefore, it is unlikely that they are associated with UDG~32. Instead, these GCs are likely intracluster GCs. Their velocities are within $\pm 1\sigma$ of the mean cluster velocity, and their number density in the area of the MUSE FoV agrees with the expected intra-cluster GC density of the Hydra~I cluster \citep{2024A&A...689A.306S, 2024A&A...683A...8G} and in the North of NGC~3314A/B \citep{2022A&A...666A..99H}. 
We discuss the implications of the absence of GCs bound to UDG~32 in Sect.~\ref{sect:discussion}.

\section{Stellar population properties of UDG 32} 
\label{sect:SSP}
We constrain the stellar population properties of UDG~32 by fitting models to the observed SED that we determine from deep imaging. The SNR of the MUSE spectra in the part of UDG~32 not covered by the filament is too low to use spectral fitting. Instead we use the Bayesian spectral fitting code \textsc{Bagpipes} \citep{2018MNRAS.480.4379C} to fit multi-band photometry of UDG~32. We fit a simple parametrisation of the SF history that is parametrised by the stellar mass, age, metallicity, and dust content of the galaxy. The detailed model setup is described in Appendix~\ref{app:SED}.  

\begin{figure}
    \centering
    \includegraphics[width=8.8cm]{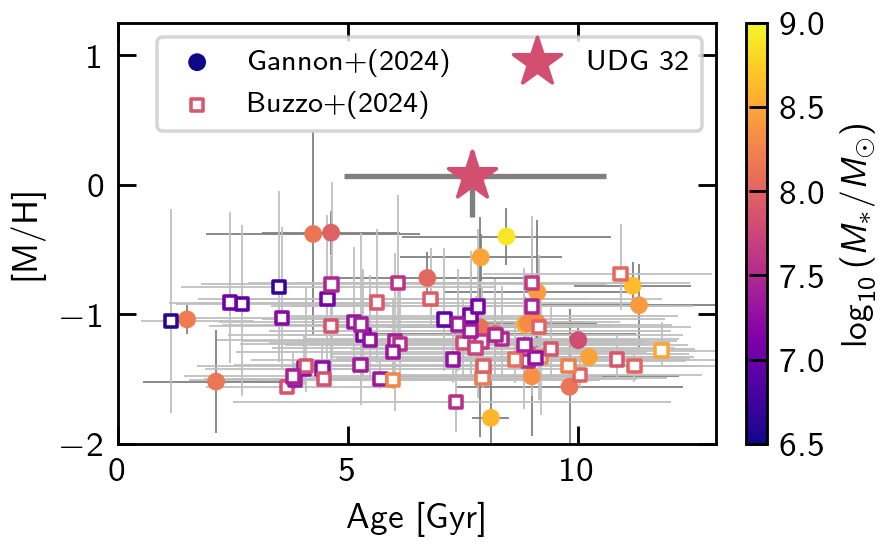}
    \caption{Mass-weighted age and metallicity of UDG~32 (star) derived via SED fitting in comparison to values for UDGs derived from spectroscopy \citep[filled circles,][]{2024MNRAS.531.1856G} and SED fitting \citep[open squares,][]{2024MNRAS.529.3210B}. The UDGs are colour-coded by their stellar mass $M_{*}$.}
    \label{fig:age-met}
\end{figure}

We fit models to the optical photometry and derive a mass-weighted age of $7.7^{+2.9}_{-2.8}\,\mathrm{Gyr}$, a metallicity of $\mathrm{[M/H]} = 0.07^{+0.19}_{-0.32}$ dex and a stellar mass of $M_{*} = 6.7^{+2.0}_{-1.8}\times 10^{7}\,M_\odot$. The stellar mass agrees well with the estimate of $M_{*}\sim 9\times10^{7}\,M_\odot$ by \citet{2021A&A...652L..11I} that is based on the integrated magnitude and colour of UDG~32.
When also including the UV data from GALEX, all fitted and derived parameters have overlapping 16\% to 84\% posterior distributions (see Table~\ref{tab:SED_results}). 
However, we cannot distinguish between UV flux from the filament and UV flux from UDG~32 due to the large full-width half maximum of the GALEX images ($\sim 5\arcsec$) and none of the best-fit models can reproduce the large observed UV fluxes.

Figure~\ref{fig:age-met} shows the stellar population properties of UDG~32 in comparison to those derived from other UDGs via spectroscopy \citep[][and references therein]{2024MNRAS.531.1856G}  and SED fitting \citep{2024MNRAS.529.3210B}. UDG~32 is more metal-rich than other UDGs in the Hydra~I cluster (Doll et al., in prep.), and both field and cluster UDGs from the literature. Compared to UDG properties derived from spectroscopy \citep{2024MNRAS.531.1856G}, UDG~32 lies at the lower end of the mass and age distribution, however, also considering stellar masses derived via SED fitting \citep{2024MNRAS.529.3210B} as in this work, UDG~32 may be considered intermediate-mass and -age. 

\section{Clues towards the formation scenario of UDG~32}
\label{sect:discussion}
\begin{table*}
    \centering
    \caption{Structural and stellar population properties predicted for UDG from formation scenarios involving gravitational interactions, compared with those for UDG~32.}
    \begin{tabular}{l|p{3.5cm}p{3.5cm}p{3.5cm}p{3.5cm}}
    \hline\hline 
    Observables & RPS clumps & Tidally heated dwarfs & TDG-like & UDG 32 \T\\
    References. & (1,2,3)& (4,5) & (6,7,8,9) & \textit{this work}; (10) \\
     \hline 
     Stellar mass & $\sim 10^{3.5} - 10^{8}\,M_\odot$ & $\sim 10^{8} - 10^{9}\,M_\odot$ & $\sim 10^{7} - 10^{9}\,M_\odot$ & $6.7^{+2.0}_{-1.8}\times 10^{7}\,M_\odot$ \T\\
     $g-r$ colour & blue & $0.4-0.5$ mag & blue & $0.54-0.61$ mag\tablefootmark{a}\T\\
     Dust content &  $A_V \lesssim 1.5$ mag& dwarf-like & $A_V \lesssim 1.5$ mag& $A_V =  0.17^{+0.26}_{-0.13}$ mag \T\\
     Age & $\sim 0.01 - 0.4$ Gyr & dwarf-like: $\sim 6-14$ Gyr & $\lesssim 4$ Gyr & $7.7^{+2.9}_{-2.8}\,\mathrm{Gyr}$ \T\\
     Metallicity \text{[M/H]} & moderate & dwarf-like (following the dwarf MZR: $\lesssim -1$ dex) & enhanced (compared to the dwarf MZR) & $0.07^{+0.19}_{-0.32}$ dex \T\\
     Gas content & yes & no & possible & no \T \\
     UV emission & yes & no & possible & to be constrained \T\\
     GC content & no & dwarf-like & no & no \T \\
     DM content & no & dwarf-like & no & to be constrained \T \\
    \hline
    \hline 
    \end{tabular}
    \label{tab:scenarios}
    \tablefoot{
    \tablefoottext{a}{Integrated colours from OmegaCAM and DECam respectively. Also note the gradient towards redder colours ($\sim 1$ mag) in the outskirts of UDG~32 reported in \citet{2021A&A...652L..11I}.}
}
    \tablebib{(1)~\citet{2019MNRAS.482.4466P}; 
              (2)~\citet{2024A&A...682A.162W};  
              (3)~\citet{2024ApJ...969...24L};
              (4)~\citet{2018ApJ...866L..11B}; 
              (5)~\citet{2021ApJ...919...72J}; 
              (6)~\citet{2012MNRAS.419...70K}; 
              (7)~\citet{2014MNRAS.440.1458D}; 
              (8)~\citet{2018MNRAS.474..580P}; 
              (9)~\citet{2024A&A...687A.105I}; 
              (10)~\citet{2021A&A...652L..11I}.}
\end{table*}
Thanks to the new MUSE data from LEWIS, in combination with deep multi-band imaging, we are able to put new constraints on the formation scenario of UDG~32. First of all, based on its LOS velocity we can now place UDG~32 into the same dynamical structure as NGC~3314A and its RPS material and associate it with the Hydra~I cluster South-East group  (Fig.~\ref{fig:vhist}). Associating UDG~32 with this group also confirms its ultra-diffuse nature according to the size definition of \citet{2015ApJ...804L..26V}. 
In Table~\ref{tab:scenarios} we present the observed properties for UDG~32. We find UDG~32 to be an intermediate-mass UDG with a red-ish colour compared to other UDGs in the Hydra~I cluster \citep{2022A&A...665A.105L}, and to have an intermediate age and near-solar metallicity (Fig.~\ref{fig:age-met}), which makes it one of the most metal-rich UDGs in the LEWIS sample\footnote{The stellar population properties of the full LEWIS sample will be presented in Doll et al. (in prep.). The ages of UDGs with reliable stellar population estimates from spectral fitting in the sample range from 2 to 12~Gyr and the metallicities from -1.5 to -0.3 dex.}. We do not detect any emission pointing towards gas associated with UDG~32. The morphology of the regions with strong emission lines is very different from that of the stellar body of UDG~32 (cf.~Sect.~\ref{sect:emission}). We also do not detect any GCs dynamically bound to the UDG, but two intracluster GCs (cf. Sect.~\ref{sect:GCs}). 

In Table~\ref{tab:scenarios}, we have also summarised the structural properties predicted by the formation scenarios involving gravitational interactions, which were already addressed by \citet{2021A&A...652L..11I} as possible formation channels for UDG~32, i.e. the formation from RPS clumps \citep{2019MNRAS.482.4466P, 2024ApJ...969...24L} stripped from NGC~3314A, the formation through tidal interaction and subsequent heating of dwarf galaxies \citep{2018ApJ...866L..11B, 2021ApJ...919...72J}, and the formation as a TDG \citep[e.g.,][]{2014MNRAS.440.1458D}. While NGC~3314A is not an actively merging galaxy, we cannot rule out that UDG~32 formed from a previous galaxy-galaxy interaction in the Hydra-I south-east group, also considering the recent simulations of \citet{2024A&A...687A.105I} showing that TDG-like objects may be longer-lived in cluster environments than previously found in simulations. 
 
The stellar mass estimated for UDG~32 is consistent with the typical values expected in all formation scenarios, ranging from $\sim 10^6\,M_\odot$ to $\sim 10^9\,M_\odot$.
UDG~32 seems to have redder colours than those expected for UDGs formed by RPS gas clumps, but the presence of dust might affect this value. UDG~32 has a ``clumpy'' morphology on the deep VST OmegaCam images, which was already indicative of the presence of dust \citep{2021A&A...652L..11I}.
The internal dust absorption of UDG~32 that we derive via SED fitting is low, but in line with the moderate dust content observed in RPS clumps with ongoing SF \citep[e.g.,][]{2024A&A...682A.162W}. Similar levels of $A_V$ have also been observed in young TDGs \citep[e.g.,][]{2012MNRAS.419...70K}. 
In UDGs resulting from the tidal heating of a dwarf galaxy interacting with a massive galaxy, dust might be present in the newly formed UDG if the original dwarf galaxy had dust. 

More constraints may come from the intermediate age and enhanced metallicity that we derive for UDG~32. Tidal interactions transforming a dwarf galaxy into a UDG would result in a galaxy with a low (dwarf-like) metallicity and older age  ($\sim 6 - 14$~Gyr). While the age of UDG~32 would be consistent with this scenario, its near-solar metallicity, points towards a formation from pre-enriched material, either liberated from more massive galaxies through RPS or during galaxy mergers (resulting in TDGs). If UDG~32 followed the dwarf galaxy mass-metallicity relation (MZR), its metallicty would be expected to be $\lesssim -1$ dex \citep{2017A&A...606A.115H}.

In both cases, these UDGs could have gas and UV emission resulting from the ongoing SF and they are devoid of DM. Conversely, UDGs formed via tidal heating are expected to have a DM content typical for dwarf galaxies (i.e., mass-to-light ratios $>> 1$), and they are gas-poor and quenched.
To constrain the DM content, it is imperative to measure the velocity dispersion of UDG~32. This is not possible with the current spectroscopic data and will require follow-up observations to either obtain deeper MUSE data improving the SNR or dedicated observations with higher spectral resolution. In \citet{2023A&A...679A..69I} and \citet{Buttitta2025}, the LEWIS collaboration has demonstrated that it is possible to recover velocity dispersion measurements below the instrumental velocity resolution of MUSE with high-quality spectra (SNR $\geq 15$) in the optical wavelength range ($4800 \leq \lambda_\mathrm{rest} \leq 7000$ \AA). 

To better constrain the gas content in UDG~32, we require UV data with higher spatial resolution than the GALEX FUV and NUV imaging with a full-width half maximum of a few arcseconds. 
Lastly, UDG~32 does not host evolved stellar systems such as GCs, which may be expected had it formed through tidal heating, unless the GCs had already been tidally stripped \citep{2023MNRAS.525L..93F}. However, the light distribution of UDG~32 is well fit by a single Sersic profile and does not show morphological asymmetries that point towards tidal disturbances \citep{2021A&A...652L..11I}. No GCs are expected in TDG-like UDGs or those from RPS material. 

Considering the unprecedented extent of the stripped material that we discovered based on the MUSE data that shows signatures of ram-pressure triggered SF (cf. Sect.~\ref{sect:emission}) and the proximity to UDG~32 in phase space, the scenario that UDG~32 formed from stripped material remains viable. However, had UDG~32 formed from one of these stripped gas clumps, its age should be a fraction of a gigayear. This is much lower than our current age estimate of several gigayears from SED fitting, even when accounting for a moderate UV emission from the galaxy and the broad posterior distribution of the mass-weighted age. The simulations of \citet{2024A&A...687A.105I} show that TDG-like DM free galaxies in massive cluster environments can be longer-lived and distributed throughout the cluster, but do not make predictions about their stellar populations. 

The Hydra~I cluster and the south-east group, in which UDG~32 resides, are dynamically rich environments, further pointing towards a formation of UDG~32 from pre-enriched material that was stripped from a more massive host galaxy, be it via ram-pressure or tidal stripping. Higher- SNR and -resolution spectroscopy to determine the galaxy's internal kinematics and matter content and spatially resolved UV data will allow us to even better constrain the nature and evolutionary history of UDG~32 in the future. 

\begin{acknowledgements}
The authors wish to thank L. Buzzo, L. Coccato, E. Congiu, E. Emsellem, A. Ferré-Mateu, C. Mondal, O. Gerhard, J. Gannon, F. Marleau, T. Puzia, and A. Werle for the useful comments and discussions on the work presented in this paper. 
We thank the anonymous referee for the careful reading of the manuscript and constructive comments. 
We thank the ESO Paranal Observatory staff for carrying out the observations used in this publication in service mode.   
J.H. and E.I. acknowledge the financial support from the visitor and mobility programme of the Finnish Centre for Astronomy with ESO (FINCA), funded by the Academy of Finland grant nr 306531. J.H. wishes to acknowledge CSC – IT Center for Science, Finland, for computational resources. 
This work is based on the funding from the INAF through the GO large grant in 2022, to support the LEWIS data reduction and analysis (PI E. Iodice). E.I. and M.C. acknowledge the support by the Italian Ministry for Education, University and Research (MIUR) grant PRIN 2022 2022383WFT “SUNRISE”,
CUP C53D23000850006.
G.D. acknowledges support by UKRI-STFC grants: ST/T003081/1 and ST/X001857/1.
C.C.d.l.C. acknowledges financial support from the grant CEX2021-001131-S funded by MICIU/AEI/ 10.13039/501100011033, from the grant PID2021-123930OB-C21 funded by MICIU/AEI/ 10.13039/501100011033, by ERDF/EU and from the grant TED2021-130231B-I00 funded by MICIU/AEI/ 10.13039/501100011033 and by the European Union NextGenerationEU/PRTR.
R.K. gratefully acknowledges financial support from the National Science Foundation under Grant No. AST-2150222.
J.F.-B. acknowledges support from the PID2022-140869NB-I00 grant from the Spanish Ministry of Science and Innovation.
D.F. thanks the ARC for support via DP220101863 and DP200102574.
EMC is supported by the Istituto Nazionale di Astrofisica (INAF) grant Progetto di Ricerca di Interesse Nazionale (PRIN) 2022 C53D23000850006 and Padua University grants Dotazione Ordinaria Ricerca (DOR) 2020-2022.

Based on observations collected at the European Organisation for Astronomical Research in the Southern Hemisphere under ESO programmes 108.222P and 099.B-0560. Based on observations made with the NASA Galaxy Evolution Explorer. GALEX was operated for NASA by the California Institute of Technology under NASA contract NAS5-98034.  

This research made use of the following open-source software packages: 
\textsc{astropy} \citep{2013A&A...558A..33A, 2018AJ....156..123A},
\textsc{astroquery} \citep{2019AJ....157...98G},  
\textsc{bagpipes} \citep{2018MNRAS.480.4379C},
\textsc{corner} \citep{2016JOSS....1...24F}, 
\textsc{Der\_Snr} \citep{2008ASPC..394..505S},
\textsc{emcee} \citep{2013PASP..125..306F}, 
\textsc{matplotlib} \citep{2007CSE.....9...90H}, 
\textsc{mpdaf} \citep{2016ascl.soft11003B, 2017arXiv171003554P},
\textsc{NebulaBayes} \citep{2018ApJ...856...89T},
\textsc{numpy} \citep{2011CSE....13b..22V},
\textsc{photutils} \citep{larry_bradley_2023_7946442},
and
\textsc{pPXF} \citep{Cappellari2023}.
This research has made use of the Astrophysics Data System, funded by NASA under Cooperative Agreement 80NSSC21M00561.

We thank \citet{2024MNRAS.531.1856G} for the compilation of their catalogue of UDG spectroscopic properties. The catalogue includes data from: \citet{mcconnachie2012, 2015ApJ...804L..26V, Beasley2016, Martin2016, 2016ApJS..225...11Y, MartinezDelgado2016, vanDokkum2016, 2017ApJ...844L..11V, Karachentsev2017, 2018Natur.555..629V, 2018ApJ...856L..31T, 2018ApJ...859...37G, 2018ApJ...862...82L, 2018MNRAS.478.2034R, Alabi2018, 2018MNRAS.479.4891F, Forbes2018, 2019MNRAS.484.3425M, 2019ApJ...884...79C, 2019A&A...625A..77F, Danieli2019, 2019ApJ...880...91V, torrealba2019, 2020A&A...642A..48I, Collins2020, 2020A&A...640A.106M, Gannon2020, 2020ApJ...899...69L, Muller2021, 2021MNRAS.500.1279F, Shen2021, Ji2021, Huang2021, 2021MNRAS.502.3144G, Gannon2022, Mihos2022, Danieli2022, Villaume2022, 2022MNRAS.516.3318W, Saifollahi2022, Janssens2022, 2023MNRAS.518.3653G, FerreMateu2023, Toloba2023, 2023A&A...679A..69I, 2023ApJ...957....6S}. 
\end{acknowledgements}

\bibliographystyle{aa_link} 
\bibliography{lewis}

\begin{appendix}
\section{Modifications of the standard data reduction workflow}
\label{app:datared}

\begin{figure}
    \centering
    \includegraphics[width=5.4cm]{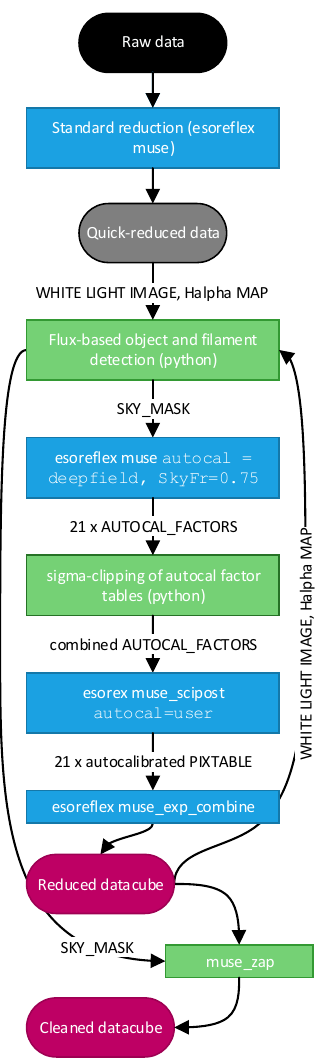}
    \caption{Data reduction flowchart sketching the process of obtaining the cleaned datacube from raw data. ESO reflex and esorex pipeline processes are shown in blue and auxiliary python routines in green boxes.}
    \label{fig:flowchart}
\end{figure}

The data are reduced with the MUSE data reduction pipeline \citep[version 2.8.4;][]{2020A&A...641A..28W} in the ESO Reflex environment \citep{2013A&A...559A..96F} and additional python scripts to create ancillary input files as shown in the flowchart in Fig.~\ref{fig:flowchart}. As for the other LEWIS UDGs, we first carry out a quick processing with the standard pipeline parameters set in the ESO Reflex MUSE canvas. As the LEWIS programme does not include dedicated sky observations for the sky subtraction, using \texttt{SkyMethod=auto}, the sky was estimated from the faintest 20\% of the image (\texttt{SkyFr=0.2}). For the exposure alignment, we set \texttt{threshold=5}, \texttt{bkgfraction=0.3}, and \texttt{srcmin=5}.

\begin{figure}
    \centering
    \includegraphics[width=8.8cm]{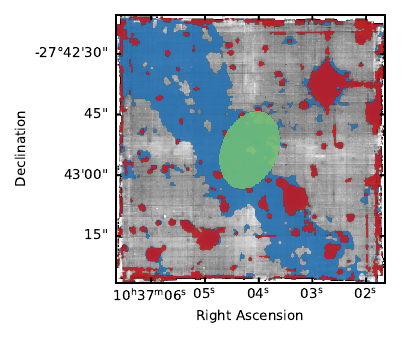}
    \caption{White-light image from the standard reduction with the H$\alpha$ mask shown in blue, the object mask in red, and the UDG mask in green. }
    \label{fig:mask_combo}
\end{figure}

\begin{figure*}
    \centering
    \includegraphics[width=18cm]{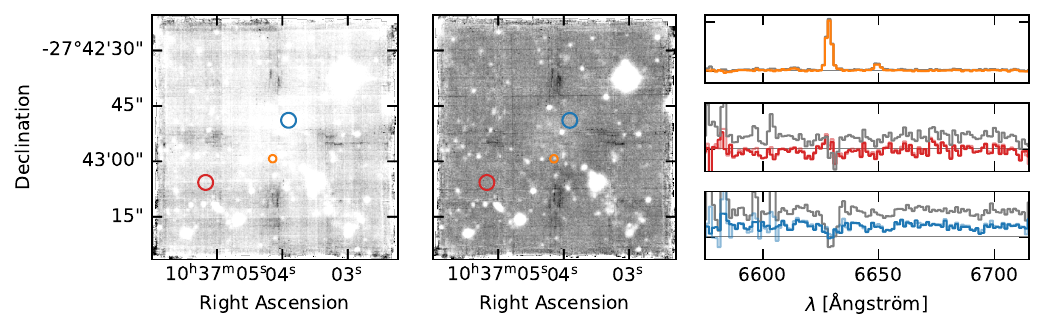}
    \caption{Comparison of different data reduction runs. The left panel shows the $r$-band image derived from the quick-reduced data and the middle one the $r$-band image derived from the improved reduction discussed in Appendix~\ref{app:datared}. The circles denote three apertures on a patch of presumably empty sky (red), a \ion{H}{ii} knot (orange), and on the stellar body of UDG~32 away from the gaseous filament (blue). The right panels correspond to the spectra from these apertures around the H$\alpha$ line with the same colour-coding. In each panel, the grey lines correspond to the spectra extracted from the quick-reduced data, the lightly coloured lines to the improved reduction before running ZAP and the opaque coloured lines to the ZAPped spectra.}
    \label{fig:redcomp}
\end{figure*}
This results in the first combined data cube that we then used to create adequate masks for a refined exposure combination and subsequent sky subtraction. Due to the extent of the gas filament (see Sect.~\ref{sect:emission}), it is paramount to mask it adequately; else the \texttt{muse\_create\_sky} workflow misinterprets the redshifted H$\alpha$ line at $\sim6630$\,\AA\ with two OH lines at $(6628,6634)\,$\AA, leading to "fake" absorption lines around this wavelength range that could lead to a wrong estimation of the redshift of UDG~32.
We construct a pseudo-narrow band image covering a 20\,\AA\ wavelength range centred on $\lambda_c= 6630$\,\AA\ and use this to mask all the pixels with strong H$\alpha$ emission. The resulting mask is shown in blue in Fig.~\ref{fig:mask_combo}. In addition to this mask, similar to the procedure for the other LEWIS UDGs \citep{2023A&A...679A..69I}, we also mask foreground and background objects (red mask in Fig.~\ref{fig:mask_combo}) as well as the central $\sfrac{1}{3}\,R_\mathrm{eff}$ of the UDG (green ellipse in Fig.~\ref{fig:mask_combo}) This ellipse has a slightly different orientation compared to the isophotes of the UDG presented in \citet{2021A&A...652L..11I}. The orientation and extent of the ellipse was determined based on the appearance of the diffuse light from the UDG after the first reduction steps. 

We use the same modified esoreflex workflow as for the other UDGs with one notable exception: as roughly 50\% of the FoV are now masked, we combine the 21 individual \texttt{AUTOCAL\_FACTORS} tables produced by running the pipeline with the \texttt{autocal=deepfield} keyword to a single one with the $\sigma$-clipping algorithm provided by the \textsc{mpdaf} package as recommended by \citet[][Sect.~3.10.2]{2020A&A...641A..28W}. The merged \texttt{AUTOCAL\_FACTORS} are then applied to the 21 \texttt{PIXTABLES} setting \texttt{autocal=user}, resulting in 21 autocalibrated \texttt{PIXTABLE\_REDUCED} data products that are combined with the modified esoreflex \texttt{muse\_exp\_combine} workflow as described in \citet{2023A&A...679A..69I}, resulting in the stacked and flux-calibrated data cube. 

To further remove residual sky contamination, we use the python implementation of the Zurich Atmospheric Purge algorithm \citep[ZAP,][]{2016MNRAS.458.3210S}. Before running ZAP, we updated the sky mask (i.e. the combination of UDG, object, and H$\alpha$ masks) based on the white light image and H$\alpha$ narrow-band image derived from the newly reduced data cube. We use the default continuum filter width (\texttt{cfwidthSP=300}) and the best-fit number of eigenspectra \texttt{nevals=91}. We also re-run ZAP for a grid of \texttt{cfwidthSP=[30,50]} and \texttt{nevals=[30,60,90]} following the best practices described in \citet{2016MNRAS.458.3210S}, but find that the larger continuum filter width combined with a higher number of eigenspectra best retains the spectral shape of the filament even over large spatial scales, while on small spatial scales and in the optical wavelength range ($\lambda \leq 8000$) the differences between the different ZAP runs is small.

To illustrate the improved data quality following the reduction above, we compare the $r$-band image derived from  the final data cube with that derived from the quick-reduced data in Fig.~\ref{fig:redcomp}. We also mark three apertures from which we extract spectra that are displayed in the right column: a \ion{H}{ii} clump (orange), an empty region of sky, and a part of the UDG that is not covered by the filament (blue). 

\section{Ancillary imaging data}
\label{app:ancdata}
In this appendix, we describe the imaging data used for the stellar population analysis in Sect.~\ref{sect:SSP} via SED fitting (detailed in Appendix~\ref{app:SED}). We focus on optical and UV data, as there is no notable signal detected at the position of UDG~32 in any of the WISE or Spitzer IRAC bands \citep[for the latter see also Fig.~B.1 in][]{2022A&A...668A.184H}. 

\subsection{Optical imaging from OmegaCAM and DECam}
UDG~32 was first detected from deep OmegaCAM \citep{1998Msngr..93...30A, 2002Msngr.110...15K} imaging in the scope of the VEGAS survey with the VST \citep{2021A&A...652L..11I} in the optical $g$- and $r$-bands with a central surface brightness $\mu_0 = 26\pm1\,\mathrm{mag}\,\mathrm{arcsec}^{-1}$, an integrated colour of $g-r = 0.54 \pm 0.14\,\mathrm{mag}$, but redder colours in its outskirts, and an absolute $r$-band magnitude of $M_r=-14.65\,\mathrm{mag}$. 
In addition to these data, the Hydra~I cluster was also surveyed with the Dark Energy Camera \citep[DECam;][]{2015AJ....150..150F} in six bands ($ugriz$ and H$\alpha$: N662). These data (PI: R. Kotulla) were first presented in \citet{2022A&A...668A.184H} and the summary of their resolution and photometric depth are given in Table~1. For this paper, we focus on the optical $ugri$ images, as both the $z$-band and narrow-band H$\alpha$ images have strong variations of their background level, which are of the same order as the UDG central flux. 

We follow the methodology in \citet{2021A&A...652L..11I} to derive the magnitudes in the $5$ broad-band and narrow-band H$\alpha$ observations, i.e. first masking foreground and background objects (including the two GCs that are projected on top of the UDG) and then deriving magnitudes in a circular aperture of 70 pixels (corresponding to 18\farcs71), corresponding to the limiting radius where the galaxy light blends into the sky fluctuations. The sky background is derived outside of that radius. We correct the magnitudes for Galactic foreground extinction \citep{1998ApJ...500..525S}. The resulting derived magnitudes are summarised in Table~\ref{tab:mags}. The colour $g-r = 0.61 \pm 0.07$ of UDG~32 derived from the DECam data is consistent with that reported by \citet[][see above]{2021A&A...652L..11I}.  

\begin{table}
    \centering
    \caption{Magnitudes of UDG~32 in several bands, derived in a circular aperture of 18\farcs71, corrected for foreground extinction }
    \begin{tabular}{lllll}
       \hline
       \hline
       Band  & $\mathrm{mag}$ & $\mathrm{mag}_\mathrm{err}$ & Instrument & Ref.\B\\
       \hline
        $u$ & 22.31 & 0.23 & DECam & (1)\\
        $g$ & 19.68 & 0.05 & DECam & (1) \\
        $r$ & 19.15 & 0.04 & DECam & (1) \\
        $i$ & 19.00 & 0.04 & DECam & (1) \\
       \hline 
       FUV & 19.25 & 0.10 & GALEX & (2,3) \T \\
       NUV & 18.87 & 0.05 & GALEX  & (2,3)\\
       \hline
       \hline
       \end{tabular}
    \tablebib{(1)~\citet{2022A&A...668A.184H}; (2)~\citet{2005ApJ...619L...1M}; (3)~\citet{2007ApJS..173..682M}.}
    \label{tab:mags}
\end{table}

\subsection{UV imaging}
UDG~32 is not detected in archival GALEX \citep{2005ApJ...619L...1M, 2007ApJS..173..682M} FUV or NUV imaging that covered the Hydra~I cluster as part of the GALEX All-Sky Imaging Survey. However, \citet{2021A&A...652L..11I} report UV emission associated with the filaments of NGC~3314A out to ~3\farcm7 from its centre, including the footprint of UDG~32 (see Fig.~A.1 therein). It is unclear whether this emission can also be associated with UDG~32, or only with the filaments. We derive FUV and NUV magnitudes from the GALEX data following the same prescription as for the optical data (see above), and report the values in Table~\ref{tab:mags}. We caution that these values should be treated as lower (upper) limits on the magnitude (flux) of UDG~32 in the UV. 

\section{Characterisation of the \ion{H}{ii} knots}
\label{app:abundances}

\begin{figure}
    \centering
    \includegraphics[width=8.8cm]{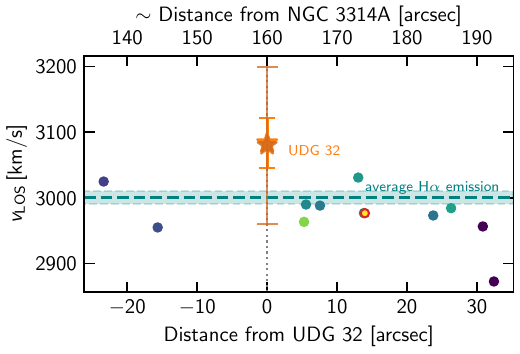}
    \caption{LOS velocity of the \ion{H}{ii} knots in the stripped material of NGC~3314A fitted with \textsc{pPXF} as function of their distance from UDG~32, with the same colour-coding as Fig.~\ref{fig:BPT}. The point circled in red denotes a knot whose spectrum is likely contaminated by a background galaxy. The teal dashed horizontal line denotes the average velocity of the stripped material and the light orange star the velocity of UDG~32 derived from the short spectrum around H$\alpha$. The darker symbol and error bars denote the conservative velocity estimate from the optical spectrum. For convenience, the top axis shows the approximate distance to NGC~3314A.}
    \label{fig:vel_distr}
\end{figure}

\begin{figure}
    \centering
    \includegraphics[width=8.8cm]{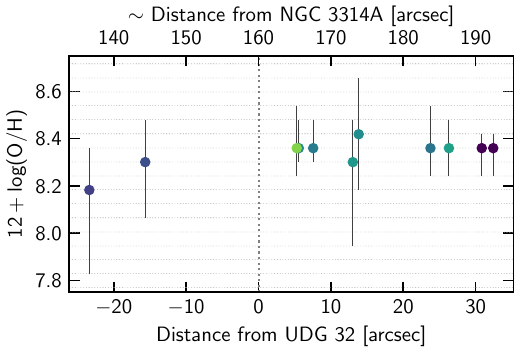}
    \caption{Abundance distribution of the \ion{H}{ii} knots in the stripped material of NGC~3314A derived with \textsc{NebulaBayes} as function of their distance from UDG~32, with the same colour-coding as Fig.~\ref{fig:BPT}. The dotted horizontal lines denote the model grid values. For convenience, the top axis shows the approximate distance to NGC~3314A.} 
    \label{fig:O/H_distr}
\end{figure}

\begin{figure}
    \centering
    \includegraphics[width=8.8cm]{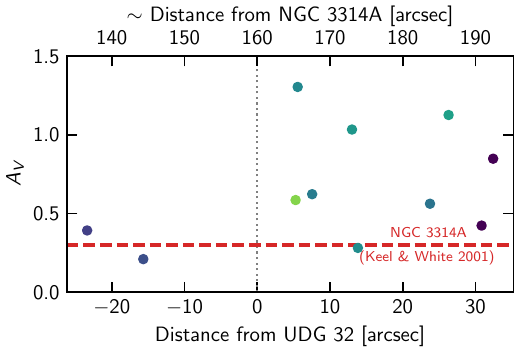}
    \caption{Extinction of the \ion{H}{ii} knots in the stripped material of NGC~3314A derived from the best \textsc{NebulaBayes} fit as function of their distance from UDG~32, with the same colour-coding as Fig.~\ref{fig:BPT}. The red dashed horizontal line denotes the typical extinction values in the interspersed dusty arms of NGC~3314A \citep{2001AJ....122.1369K}. For convenience, the top axis shows the approximate distance to NGC~3314A.}
    \label{fig:Av_distr}
\end{figure}

In Sect.~\ref{sect:emission}, we classified 11 \ion{H}{ii} knots  as star-forming based on their ELRs (cf. Fig.~\ref{fig:BPT}). We compare their emission-line fluxes to three-dimensional model grids from  \textsc{Mappings} 5.1 \citep{2017ApJS..229...34S} for \ion{H}{ii} regions to constrain their oxygen abundance ($12 + \log\mathrm{O/H}$), ionisation parameter ($\log U$, i.e. the ratio of ionizing photon to hydrogen densities) and
gas pressure ($\log P/k$) with the \textsc{NebulaBayes} package \citep{2018ApJ...856...89T}. 

Figure~\ref{fig:vel_distr} shows the velocity distribution of the knots from the \textsc{pPXF} emission-line fits that are the basis for the follow-up analysis with \textsc{NebulaBayes}. Figure~\ref{fig:O/H_distr} shows the abundance distribution of the \ion{H}{ii} knots in the stripped arm of NGC~3314A versus their distance from the centre of UDG~32 and Figure~\ref{fig:Av_distr} the corresponding extinction values of the best fits. There is a large scatter in the latter, similar to what is found by \citet{2024A&A...682A.162W} in the star-forming tails of jellyfish galaxies. The $\ion{H}{ii}$ knots have a mean ionization parameter $\left<\log U\right>= -3.3\pm0.1\,\mathrm{dex}$ and a mean pressure of $\left< P/k\right>= 371\pm62\,\mathrm{cm}^{-3}\,\mathrm{K}$.

\section{Deriving the stellar kinematics of UDG~32}
\begin{figure*}
    \centering
    \includegraphics[width=18cm]{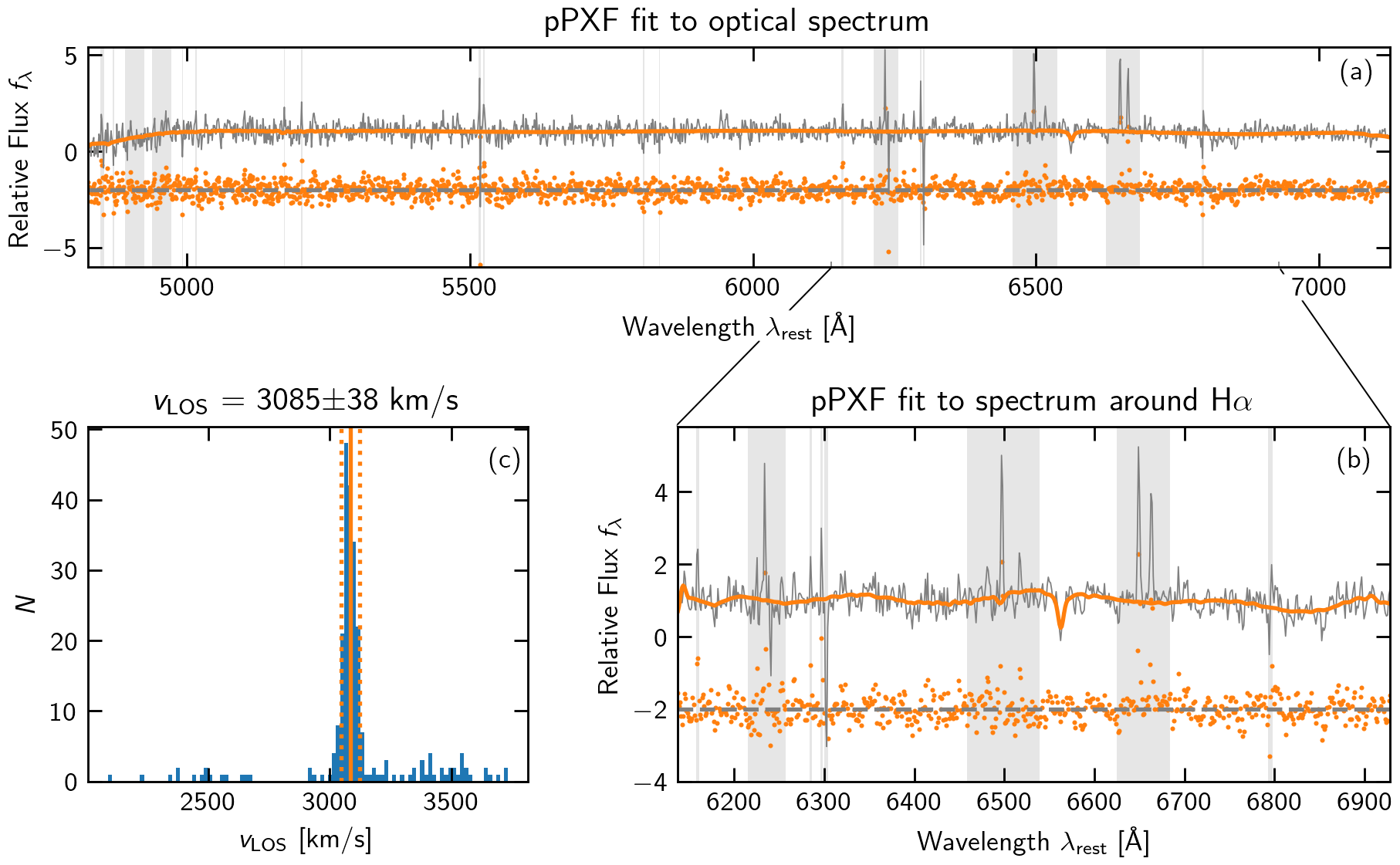}
    \caption{\textit{a:} Optical spectrum of UDG~32 extracted from the 3\arcsec-radius aperture in Fig.~\ref{fig:overview} (grey) with the best fit from \textsc{pPXF} overplotted in orange. Light grey vertical bands denote masked regions. 
    \textit{b:} Same as \textit{panel~a}, but restricted to the region around the H$\alpha$ line with a separate \textsc{pPXF} fit. 
    \textit{c:} Posterior probability distribution of $v_\mathrm{LOS}$ corresponding to the wild bootstrapping of the \textsc{pPXF} fit shown in \textit{panel~b}. The orange vertical lines show the median velocity (solid) and its uncertainties (dotted).
    }
    \label{fig:ppxf}
\end{figure*}

Figure~\ref{fig:ppxf}a shows the optical spectrum of UDG~32 extracted from a 3\arcsec-radius aperture that maximises the radius of spatial coverage while avoiding regions with strong emission-line signatures of the filament (cf. Sect~\ref{sect:emission}) and the two GCs from which we aim to derive the stellar kinematics of UDG~32. Considering the low SNR of the spectrum in this aperture \footnote{The input optical spectrum (after masking the foreground ISM emission line and iteratively clipping sky lines and other outliers following the first \textsc{pPXF} fit) has $\mathrm{SNR}\approx 3$.}, we restrict our analysis to the first two moments of the line-of-sight velocity distribution (LOSVD), i.e. the LOS velocity $v_\mathrm{LOS}$ and the velocity dispersion $\sigma_\mathrm{LOS}$, and we only fit the optical spectrum up to $\lambda_\mathrm{obs} \leq 7100\,\r{A}$. 

Running \textsc{pPXF} in conjunction with the \textsc{E-Miles} stellar library on the optical spectrum and after iteratively masking outliers, we determine a LOS velocity of $v_\mathrm{LOS} = 3080\pm120\,\mathrm{km}\,\mathrm{s}^{-1}$. We estimate the uncertainties by repeating the fit 250 times for bootstrapped realisations of the spectrum \citep[``wild bootstrapping'';][]{2018MNRAS.480.1973K, 2019A&A...625A..76E}  and also allowing for a 10\% variation of the initial values for $v_\mathrm{LOS}$ and $\sigma$. The spectrum has little constraining power bluewards of $\lambda_\mathrm{rest} \leq 5000\,\r{A}$, i.e. around the H$\beta$ line (see Fig.~\ref{fig:ppxf}a) and we do not detect any other strong absorption lines in this wavelength range. We therefore re-run ppxf with the same setup on a spectrum with narrower wavelength range ($6200 \leq \lambda_\mathrm{obs} \leq 7000\,\r{A}$) and slightly higher SNR ($\approx 3.5$). The resulting spectrum and fit are shown in Fig.~\ref{fig:ppxf}b and we determine a LOS velocity of $v_\mathrm{LOS} = 3085\pm38\,\mathrm{km}\,\mathrm{s}^{-1}$. The uncertainties are again estimated by wild bootstrapping and the posterior distribution of the parameters is shown in Fig.~\ref{fig:ppxf}c. As a sanity check, we repeat this analysis on spectra that were extracted before running ZAP on the cube and find LOS velocities that agree within the errors with the values reported previously in this section: $v_\mathrm{LOS} = 3084\pm165\,\mathrm{km}\,\mathrm{s}^{-1}$ for the optical spectrum and $v_\mathrm{LOS} = 3088\pm37\,\mathrm{km}\,\mathrm{s}^{-1}$ for the shorter spectrum around H$\alpha$. 

\label{app:kinematics}

\section{Globular clusters in the MUSE FoV}
\label{app:GCs}

\begin{figure}[t]
    \includegraphics[width=8.8cm]{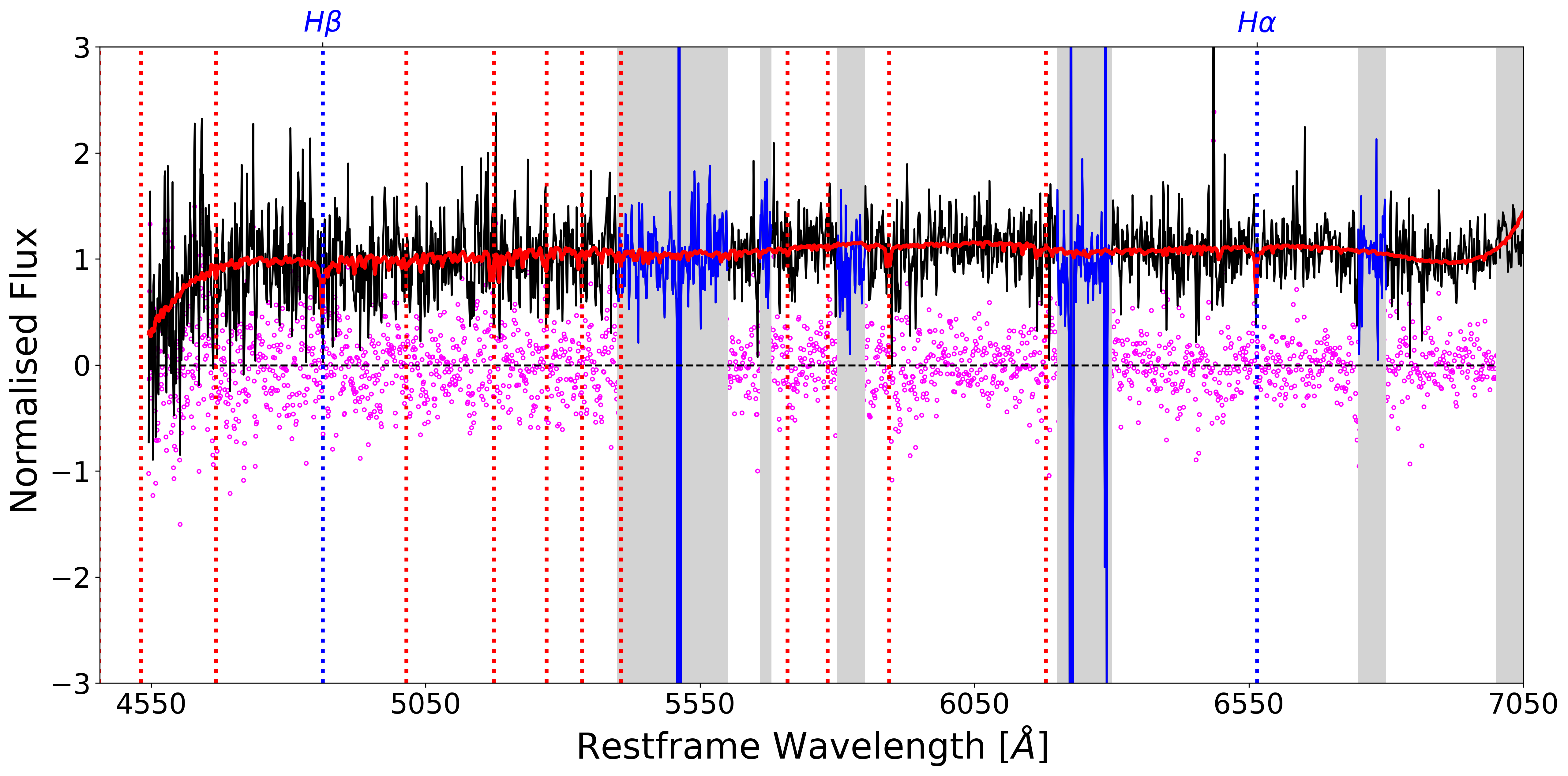}
    \includegraphics[width=8.8cm]{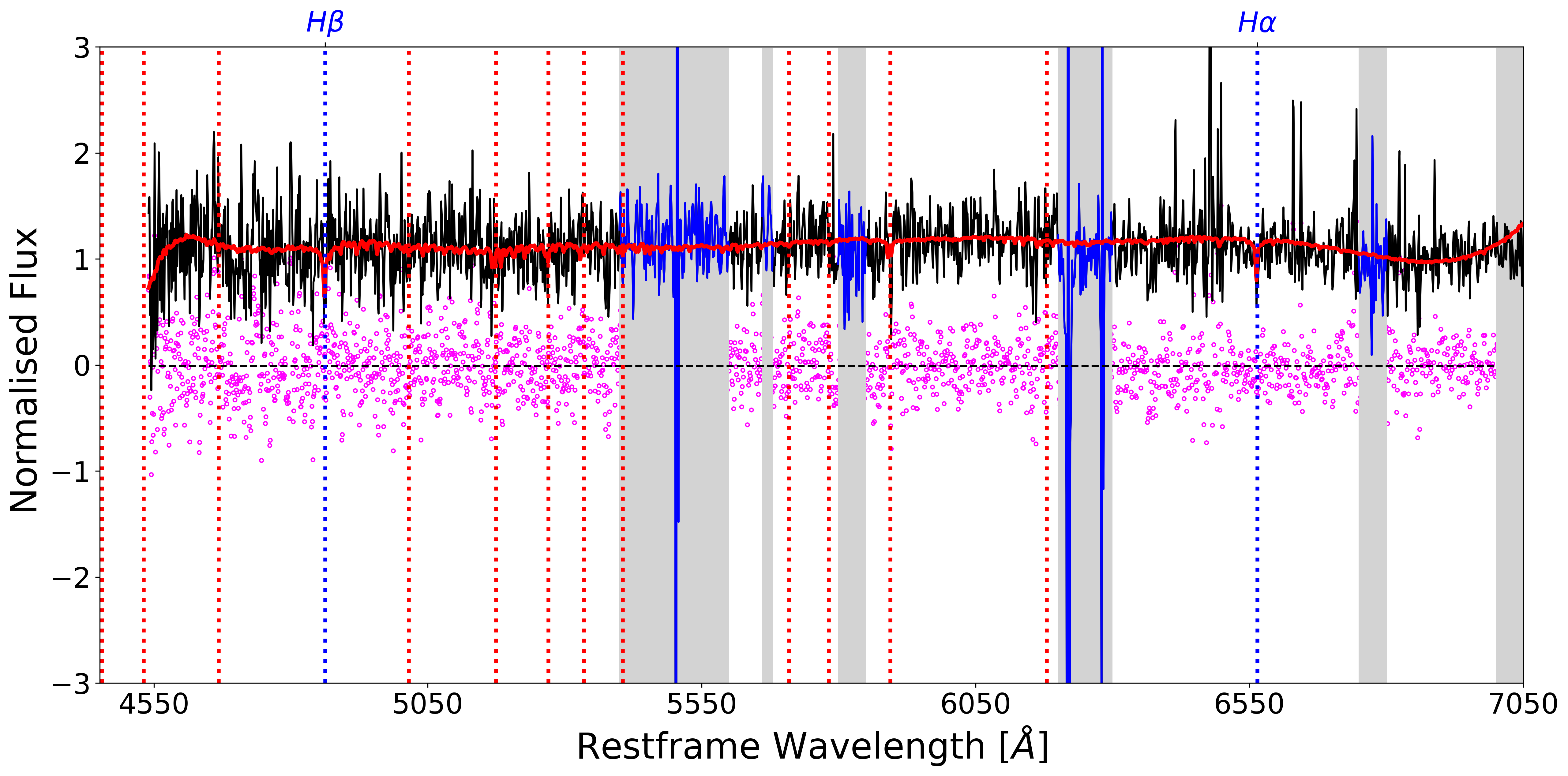}
    \caption{Extracted spectra (black) and \textsc{pPXF} fits (red) for GC1 \textit{(top)} and GC2 \textit{(bottom)}. Gray lines and regions denote the masked parts of the spectrum. The dotted vertical red lines denote the location of common absorption lines. We highlight the location of the H$\alpha$ and H$\beta$ absorption lines with the dotted vertical blue lines.}
    \label{fig:GCspec}
\end{figure}

Based on VST $g$- and $r$-band imaging, \citet{2021A&A...652L..11I} report a GC candidate density of $3 \pm 3\,\mathrm{arcmin}^{-2}$ using the methods described in \citet{2018A&A...611A..93C, 2020A&A...639A.136C} and \citet{2020A&A...642A..48I} and correcting for the contamination of foreground stars, background stars, and the presence of intra-cluster GCs. However, as shown in the LEWIS pilot paper \citep{2023A&A...679A..69I}, only two optical photometric bands are not ideal for a robust GC detection. For example, while none of the GC candidates from photometry in UDG~11 were spectroscopically confirmed, four new GCs could be identified. 

We follow the strategy of \citet[][Sect 5.5]{2023A&A...679A..69I} to extract and analyse GCs in the MUSE cube of UDG~32. We measure the LOS velocity of the candidate GCs using pPXF in conjunction with the \textsc{E-MILES} stellar library. We fit the SSP models to background-subtracted spectra extracted from 8-pixel circular apertures. However, due to the higher noise levels in the near-infrared, we only fit the optical spectra up to $\lambda_\mathrm{max,rest} = 7000 \,\r{A}$. We spectroscopically confirm two GCs, whose location is highlighted by purple circles on the rightmost panel of Fig.~\ref{fig:overview} and whose spectra are shown in Fig.~\ref{fig:GCspec}. To determine the LOS velocity uncertainties, we use the same Monte Carlo technique approach as \citet{2023A&A...679A..69I}. We report the GC properties in Table~\ref{tab:GCs}. 
We discuss the kinematic association of the GCs with the Hydra~I intracluster GC population in Sect.~\ref{sect:GCs}. 

\begin{table*}
    \centering
    \caption{Properties of the two spectroscopically confirmed GCs in the UDG~32 MUSE FoV.}
    \begin{tabular}{ccccc}
    \hline
    \hline
         ID & RA & Dec & $m_g$ & $v_\mathrm{LOS}$\\ 
          & & & [mag] & [$\mathrm{km}\,\mathrm{s}^{-1}$]\\
         \hline
         GC1 & 10:37:04.0357 & -27:42:51.939 & $23.9\pm0.1$ & $3594\pm23$ \\
         GC2 & 10:37:04.2881 & -27:42:45.283 & $24.3\pm0.1$ & $3905\pm26$ \\
    \hline
    \hline
    \end{tabular}
    \label{tab:GCs}
\end{table*}

\section{Spectral energy distribution fitting}
\label{app:SED}
\begin{figure*}
    \centering
    \includegraphics[width=18cm]{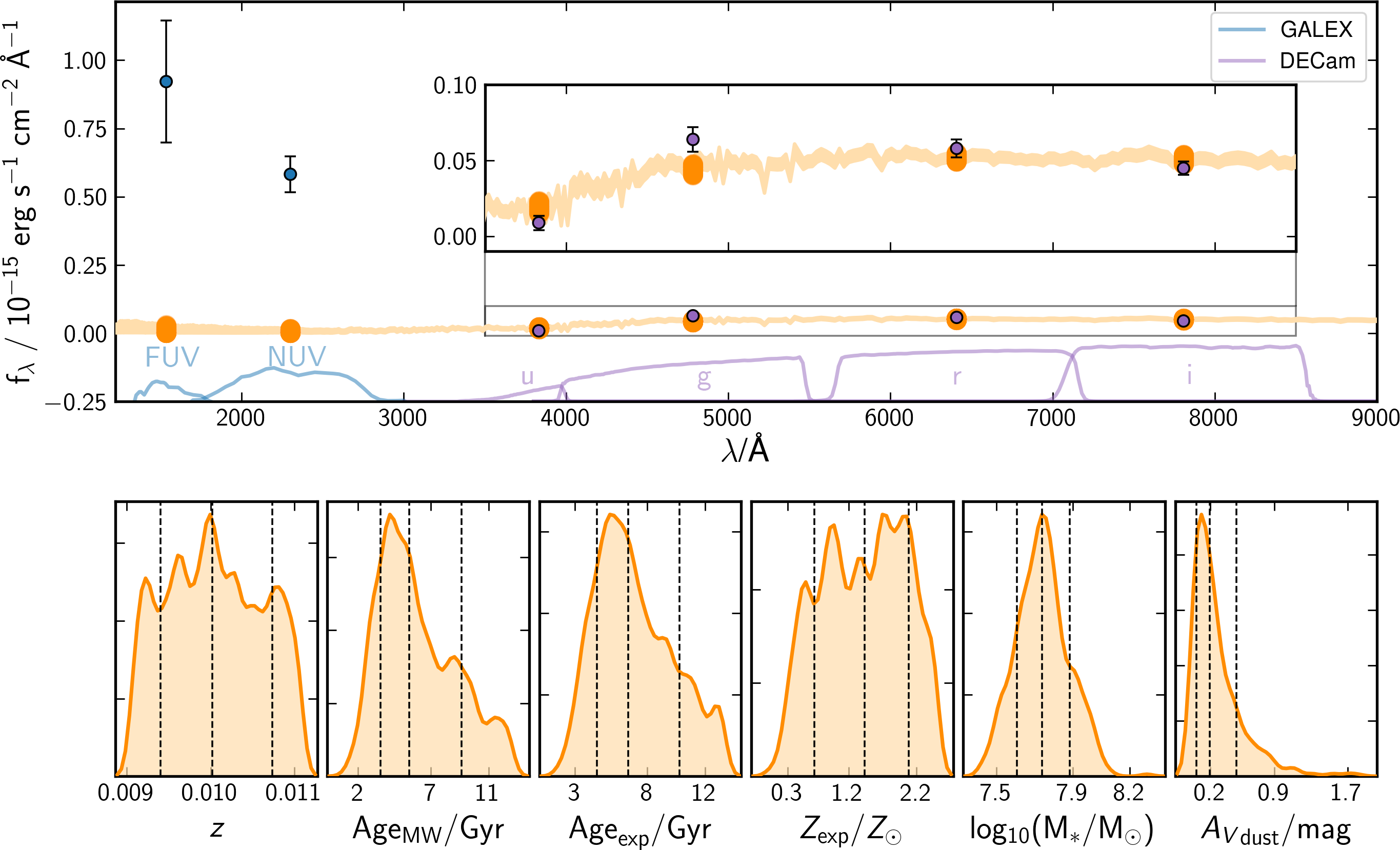}
    \caption{\textsc{Bagpipes} fitting of the UV and optical SED. \textit{Top:} Observed SED (coloured points) and best-fit model spectrum (orange). The lines at the bottom of the panel show the filter bandpasses of the imaging data used in the fit. The inset shows a zoom into the optical part of the SED. 
    \textit{Bottom:} Posterior probability distribution of the fitted and derived model parameters redshift $z$, mass-weighted age $\mathrm{Age}_\mathrm{MW}$, the time since the SFH began $\mathrm{Age}_\mathrm{exp}$, metallicity $Z_\mathrm{exp}$, stellar mass $M_{*}$, and dust extinction $A_V$. The 16, 50, and 84 percentiles are shown with vertical dashed lines.}
    \label{fig:bagpipes-all}
\end{figure*}

The MUSE data of UDG~32 that are not overlapping with the filament in projection have a prohibitively low SNR for deriving stellar population properties \citep[see also][]{2023A&A...679A..69I}, therefore, fitting the SED is a viable alternative that we explore in the following. We use \textsc{Bagpipes} \citep[Bayesian Analysis of Galaxies for Physical Inference and Parameter EStimation;][]{2018MNRAS.480.4379C} to infer stellar population parameters for UDG~32 based on the multi-wavelength imaging data presented in Appendix~\ref{app:ancdata}. \textsc{Bagpipes} uses pre-defined stellar population synthesis models \citep{2003MNRAS.344.1000B} and a \citet{2002MNRAS.336.1188K} initial mass function (IMF). We use the inbuilt exponentially declining star-formation history (SFH) parametrised by $\tau$, the timescale for the SF decrease and $T_0$, the time since the SF began. Our model furthermore includes dust attenuation according to \citet{2000ApJ...533..682C}. The inclusion of dust attenuation in the model is motivated by the work of \citet{2022MNRAS.517.2231B}, who find that the inclusion of small amounts of dust results in better agreement of stellar population properties derived from SED fitting with those from spectroscopy. We place uniform priors on the modelled redshift (with limits corresponding to the conservative uncertainties on the LOS velocity of UDG~32 estimated in Sect.~\ref{ssect:LOSV}) and the other free model parameters.  
The observed photometry and best-fit model spectrum are shown in the top panel of Fig.~\ref{fig:bagpipes-all}. The posterior distributions for the fitted and derived model parameters are shown in the bottom panels. The best-fit parameters are summarised in Table~\ref{tab:SED_results}.

\begin{table*}
    \centering
    \caption{SED fit results with \textsc{Bagpipes}. The first column shows the input data sets and the following columns the best-fit redshift $z$, the mass-weighted age, metallicity [M/H], stellar mass $M_\star$ and dust extinction $A_V$.}
    \begin{tabular}{lllllll}
    \hline\hline
    Data & $z$ & $v_\mathrm{LOS}$ & Age & [M/H] & $M_{*}$ & $A_V$ \\
     & & [$\mathrm{km}\,\mathrm{s}^{-1}$] & [Gyr] & [dex] & [$\times 10^{7}\,M_\odot$] & [mag] \\
    \hline 
    DECam $ugri$ & $0.01014^{+0.00061}_{-0.00069}$ & $3025^{+180}_{-206}$ & $7.7^{+2.9}_{-2.8}$ & $0.07^{+0.19}_{-0.32}$ & $6.7^{+2.0}_{-1.8}$ & $0.17^{+0.26}_{-0.13}$
    \T\\ 
    GALEX+DECam $ugri$ & $0.01010^{+0.00060}_{-0.00071}$ & $2984^{+211}_{-179}$ & $5.4^{+3.9}_{-2.0}$ & $0.12^{+0.20}_{-0.33}$ & $5.3^{+2.2}_{-1.5}$ & $0.20^{+0.36}_{-0.15}$ \T\B\\
    \hline\hline
    \end{tabular}
    \label{tab:SED_results}
\end{table*}

\begin{figure*}
    \centering
    \includegraphics[width=18cm]{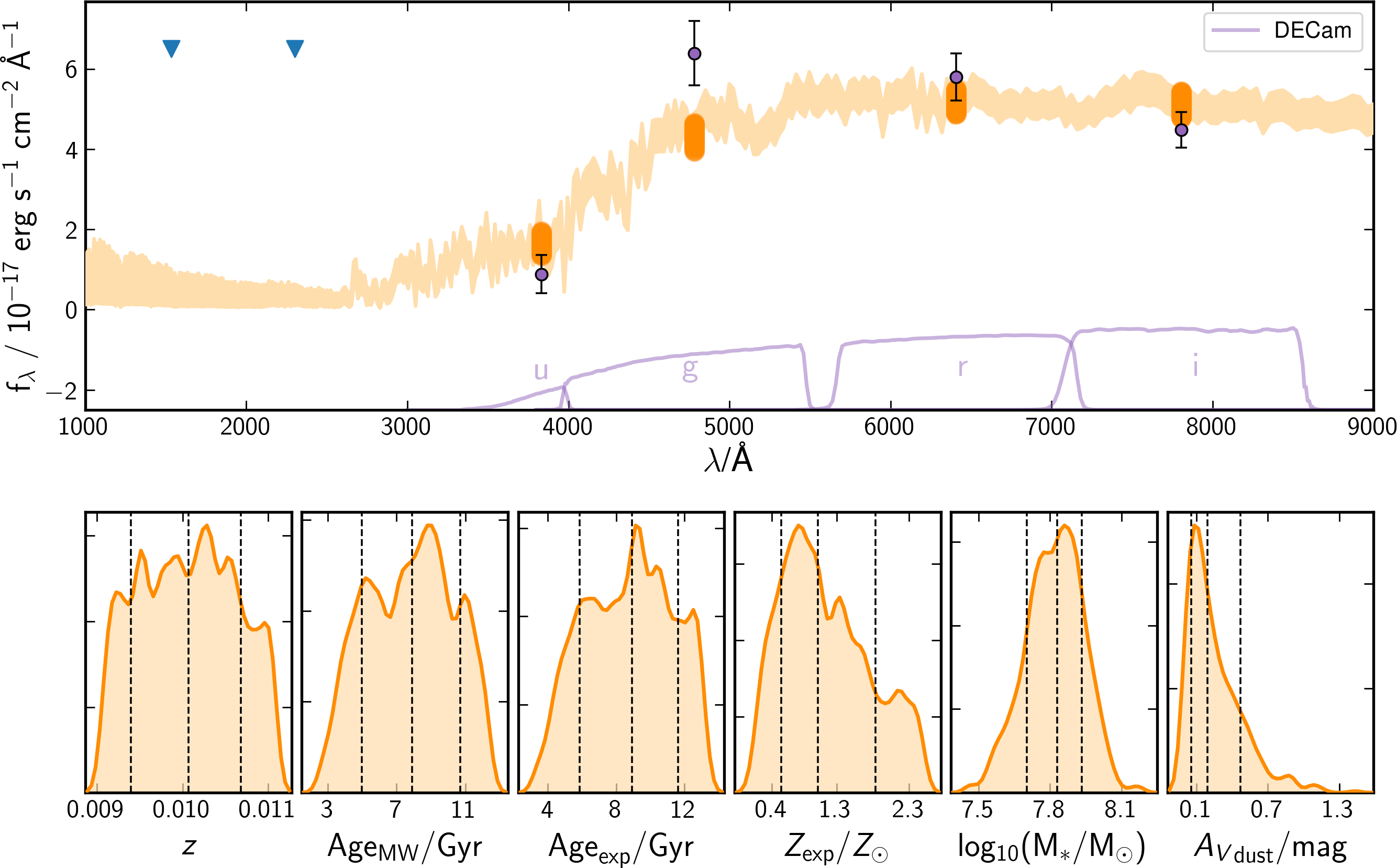}
    \caption{Same as Fig.~\ref{fig:bagpipes-all}, but only the optical data are fitted. For comparison, we show the UV data as upper limits (blue triangles) in the top panel.}
    \label{fig:bagpipes-optical}
\end{figure*}

The best-fit model cannot reproduce the high fluxes in the FUV and NUV. Considering the several-arcsecond-wide point-spread function of the GALEX data, we cannot rule out that the UV emission stems only from the filament and is not associated with UDG~32 at all. We therefore re-fit the optical photometry excluding the GALEX data points. The fit excluding the UV data is shown in Fig.~\ref{fig:bagpipes-optical}, and the best-fit parameters are presented in Table~\ref{tab:SED_results}. Both fits result in an intermediate-age population with slightly super-solar metallicities and a low dust content comparable to that of quiescent UDGs \citep{2022MNRAS.517.2231B}. The stellar mass is in the range typical for UDGs ($\sim 6\times10^{7}\,M_\odot$). 
We can use these models to predict upper limits on the integrated UV fluxes of UDG~32: 
the largest model fluxes consistent with the data are $f_\lambda = 1.03^{-0.15}_{+3.25}\times10^{-17}\,\mathrm{erg}\,\mathrm{s}^{-1}\,\mathrm{cm}^{-2}\,\mathrm{\AA}^{-1}$ in the FUV and $f_\lambda = 0.67^{-0.18}_{+1.82}\times10^{-17}\,\mathrm{erg}\,\mathrm{s}^{-1}\,\mathrm{cm}^{-2}\,\mathrm{\AA}^{-1}$ in the NUV. 

\section{Serendipitous discoveries}
\label{app:discoveries}
\subsection{Diffuse Milky Way interstellar medium}
\begin{figure*}
    \centering
    \includegraphics[width=18cm]{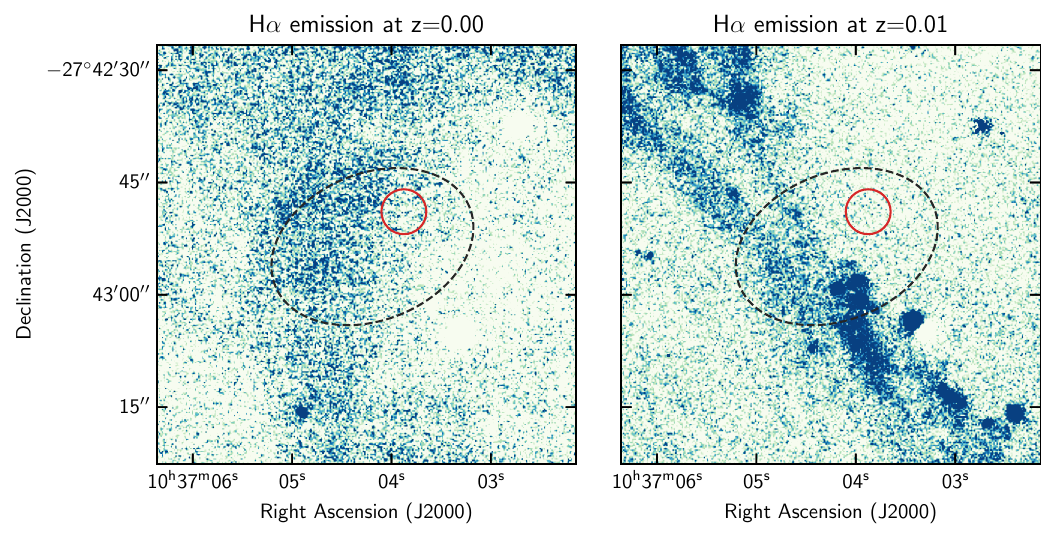}
    \caption{Maps centred on the H$\alpha$ emission at redshifts $z=0.00$ \textit{(left panel)} and $z=0.01$ \textit{(right panel)} with the same intensity cuts. For reference, we again show the aperture in which we extract the spectrum of UDG~32 in red and the footprint of the UDG from photometry \citep{2021A&A...652L..11I} with the dashed ellipse.}
    \label{fig:ISM_z0}
\end{figure*}

Upon inspection of the data cube, we also identify a diffuse emission of H$\alpha$ at redshift $z=0$. We also identify other emission lines at this redshift such as H$\beta$, [\ion{O}{iii}], [\ion{N}{ii}], and [\ion{S}{ii}] that are not sky lines. 
The left panel of Fig.~\ref{fig:ISM_z0} shows the spatial extent of this local diffuse emission in comparison to the footprint of the UDG and the emission related to the stripped material of NGC~3314A that we show in the right panel. The geometries of the two emission structures at different redshift are very distinct and therefore presumably also of distinct origin, with the foreground emission being due to the diffuse ISM of the Milky Way that happens to be along the line of sight. 

\subsection{A $z\approx0.77$ galaxy behind the filament}
\label{app:cont_galaxy}
\begin{figure}
    \centering
    \includegraphics[width=8.8cm]{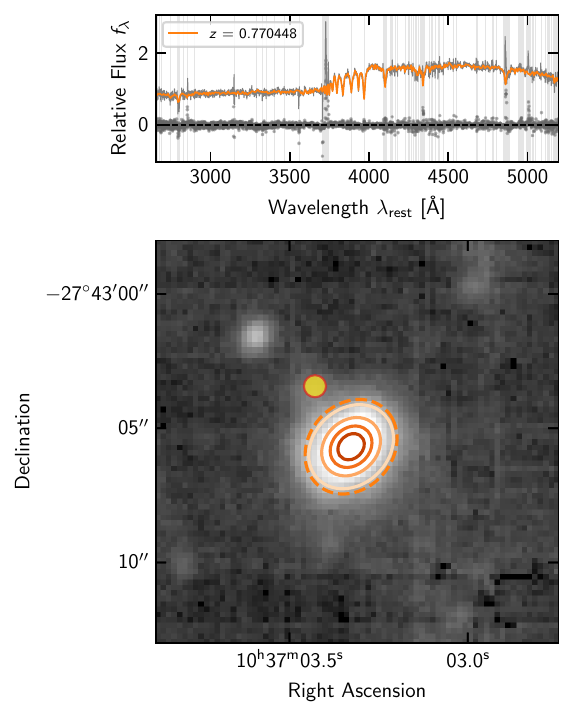}
    \caption{Background galaxy at redshift $z=0.770448$ in the southwest of UDG~32. The \textit{top panel} shows the \textsc{pPXF} fit of the absorption-line kinematics extracted from a $1~R_\mathrm{h}$ aperture and the \textit{bottom panel} the $I$-band cutout obtained from the MUSE cube to which a Moffat profile was fit to determine the effective radius. The orange dashed ellipse is the aperture from which the above spectrum was extracted. The point circled in red denotes a knot whose spectrum is likely contaminated by this background galaxy.}
    \label{fig:bkg-galaxy}
\end{figure}

In Sect.~\ref{sect:emission} we describe the stripped material from NGC~3314A and in particular the star-forming \ion{H}{ii} knots. We identify that 11/12 knots as star forming based on the theoretical demarcation between starburst and active galaxies. To investigate the nature of the one outlier (highlighted with a red outline in Fig.~\ref{fig:BPT}), we inspect the MUSE cube in its vicinity and report the serendipitous discovery of a red background galaxy 2\farcs7 to its southwest. We extract an $I$-band image from a subcube centred on the background galaxy and fit a Moffat profile to the resulting isophotes (lower panel of Fig.~\ref{fig:bkg-galaxy}). We determine a halflight radius $R_\mathrm{h} = 2\farcs03$ and an axis ratio $b/a = 1.22$. We extract a spectrum within $R_\mathrm{h}$ as shown in the top panel of Fig.~\ref{fig:bkg-galaxy} and use \textsc{pPXF} to determine the galaxies redshift and integrated dynamics based on its absorption-line kinematics. We find $z=0.770448$ and $\sigma_\mathrm{LOS} = 141\,\mathrm{km}\,\mathrm{s}^{-1}$. This galaxy is likely the object WISEA J103703.35-274306.5 that was detected in WISE bands W1, W2, and W3 in the infrared \citep{2013wise.rept....1C}, but whose redshift had not been determined previously. 

\end{appendix}
\end{document}